\documentclass{article}
\usepackage{graphicx}  
\usepackage{amsmath}   
\usepackage[compress]{cite}
\usepackage{amssymb}   
\usepackage{bm} 
\usepackage{dcolumn}
\usepackage{color}
\usepackage{mathrsfs}
\usepackage{amsfonts}
\usepackage{varioref}
\usepackage{braket}
\usepackage{subcaption}
\usepackage{comment}
\RequirePackage[colorlinks,citecolor=blue,urlcolor=magenta,linkcolor=blue]{hyperref}
\newcounter{mnotecount}[section]

\renewcommand{\themnotecount}{\thesection.\arabic{mnotecount}}

\newcommand{\mnotex}[1]
{\protect{\stepcounter{mnotecount}}$^{\mbox{\footnotesize
$
\bullet$\themnotecount}}$ \marginpar{
\raggedright\small\em
$\!\!\!\!\!\!\,\bullet$\themnotecount: #1} }

\labelformat{section}{Section #1} 
\labelformat{subsection}{Section #1} 
\labelformat{subsubsection}{Section #1}
\labelformat{subsubsubsection}{Section #1}
\labelformat{equation}{Eq.~(#1)} 
\labelformat{figure}{Fig.~#1} 
\labelformat{subfigure}{Fig.~\thefigure#1} 
\labelformat{table}{Tab.~#1} 
\labelformat{appendix}{Appendix #1}
\title{\bf Bouncing with shear: Implications from quantum cosmology}
\author{Karthik Rajeev\footnote{karthikrajeev.kr@gmail.com}$~^{1}$, Vikramaditya Mondal\footnote{vikram.iisermail@gmail.com}$~^{1}$ and Sumanta Chakraborty\footnote{sumantac.physics@gmail.com}$~^{1}$
\\
\\
{$~^{1}$\small{School of Physical Sciences}}\\
{\small{Indian Association for the Cultivation of Science, Kolkata-700032, India}}}
\begin{document}

\maketitle
\begin{abstract}
We consider the introduction of anisotropy in a class of bouncing models of cosmology. The presence of anisotropy often spells doom on bouncing models, since the energy density due to the anisotropic stress outweighs that of other matter components, as the universe contracts. Different suggestions have been made in the literature to resolve this pathology, classically. Here, we introduce a family of bouncing models, in which the shear density can be tuned to either allow or forbid classical bouncing scenarios. Following which, we show that quantum cosmological considerations can drastically change the above scenario. Most importantly, we find that quantum effects can enable a bounce, even when the anisotropic stress is large enough to forbid the same classically. We employ the solutions of the appropriate mini-superspace Wheeler-deWitt equation for homogeneous, but anisotropic cosmologies, with the boundary condition that the universe is initially contracting. Intriguingly, the solution to the Wheeler-deWitt equation exhibit an interesting phase transition-like behaviour, wherein, the probability to have a bouncing universe is precisely unity before the shear density reaches a critical value and then starts to decrease abruptly as the shear density increases further. We verified our findings using the tools of the Lorentzian quantum cosmology, along with the application of the Picard-Lefschetz theory. In particular, the semi-classical probability for bounce has been re-derived from the imaginary component of the on-shell effective action, evaluated at the complex saddle points. Implications and future directions have also been discussed.
   
\end{abstract}
\newpage
\tableofcontents
\section{Introduction}

The inflationary paradigm\cite{Guth:1980zm,Sato:1980yn,Linde:1981mu,Albrecht:1982wi,Starobinsky:1980te}, in conjunction with the standard Big-Bang cosmology, is widely considered as the most promising model of the early Universe. The proposal of a cosmic inflation gained repute by resolving several deficiencies of the Big-Bang model. However, there are also important challenges faced by the inflationary scenario (see, for instance, \cite{Borde:1993xh,Martin:2000xs,Brandenberger_2013}), which has stimulated efforts to pursue alternate paradigms for early universe cosmology. Bouncing models are one of the alternative class of early universe cosmologies, where an initially contracting universe reaches a minimum size and smoothly transitions, i.e., `bounces', into an expanding one, thereby avoiding the big bang singularity. In recent years, explorations of bouncing cosmological models have seen a significant increase in interest and widespread attention (for a small sample of works, see \cite{Novello:2008ra,Nandi:2019xag,Raveendran:2018yyh,Lilley:2015ksa,Ijjas:2015hcc,Raveendran:2017vfx,Chowdhury:2015cma,Li:2016xjb}). 

Bouncing models aim to address the flaws of the Big-Bang cosmology, \emph{without} invoking inflation. For instance, a smooth bouncing universe, by construction, is devoid of an initial singularity of the kind harboured by the Big-Bang model. Since the universe contract to a non-zero minimum size in the bouncing models, the Trans-Plankian problem can also be bypassed effortlessly (see, \cite{Brandenberger_2013}). Moreover, several of the successes of inflationary models have been claimed to be reproduced in suitably tailored bouncing scenarios, see \cite{Finelli:2001sr,Brandenberger:2012zb}. These include, for instance, the resolution of the horizon problem. Nevertheless, there are still unresolved pathologies in the bouncing universe paradigm as well. The reader may consult \cite{Brandenberger:2016vhg} for a recent review of problems and progress in these directions.   

Another set back for the bouncing models of cosmology is the existence of a potential instability in the presence of anisotropy \cite{Pereira:2007yy,Grain:2020wro}. In the standard Big-Bang cosmology the anisotropic shear is generally ignored, since its contribution to the total energy density decreases much faster than the energy densities of matter and  radiation, and hence becomes negligible as the universe expands to a large volume. However, in bouncing models one has to carefully address epochs in which the shear contribution can not be ignored. The reason being, when the universe contracts before a bounce, the increasing shear/anisotropy can potentially lead to Belinski-Khalatnikov-Lifshitz (BKL)-like instability \cite{Belinsky:1970ew}. Even in cases where instabilities do not arise, the presence of anisotropy can have other important observable consequences (see, for instance, \cite{Agullo:2020iqv}). Instability ensues, for instance, when the growing shear energy density start to overwhelm the contribution from the bounce-enabling matter field. Consequently, one deals with a tussle between the shear energy density and the bounce-enabling matter. Thus anisotropic shear must be treated in the context of bouncing models with due care.

Following the existence of classical instability for anisotropic bouncing models, in this work we will try to understand whether quantum effects can help in stabilizing the scenario. For that purpose and to illustrate the basic features, we will study the quantum aspects of the Bianchi-I bouncing models of cosmology. Taking a cue from our previous work \cite{Rajeev:2021lqk}, where we had mainly focused on the mini-superspace of spatially flat Friedmann–Lemaître–Robertson–Walker spacetimes, here we adapt the same procedure to deal with the more realistic setting of homogeneous but anisotropic bouncing spacetimes. As we will see, this will help us to understand the fate of the classical instability from the perspective of quantum cosmology.

In the literature, however, one finds several models of bouncing cosmologies, with varying physical concepts and motivations \cite{Battefeld:2014uga}. One can easily imagine that these variety of models would correspond to the variety of manners in which one can bypass the Hawking-Penrose singularity theorem \cite{Hawking:1969sw}. As expected, this may be achieved by invoking beyond standard model physics, like, for instance, modified theories of gravity \cite{Brandenberger:2012zb,Brandenberger:2009yt,Bamba:2013fha,Desai:2015haa}, exotic matter\cite{ArkaniHamed:2003uy,Cai:2007qw,Cai:2008qw,Raveendran:2017vfx,Raveendran:2018why}, asymptotic safe gravity \cite{Bonanno:2017gji,Platania:2020lqb,Platania:2019qvo}, approaches to quantum gravity/cosmology \cite{Bamba:2014zoa,Basile:2021amb,Brandenberger:1988aj,Haro:2015oqa,Ashtekar:2008ay,WilsonEwing:2012pu,Cai:2014zga}, etc. Here, as in our previous work \cite{Rajeev:2021lqk}, we adopt a phenomenological approach, that covers a wide range of bouncing scenarios without diving into the details of what unconventional physics gives rise to such scenarios. 

The paper is organized as follows --- We begin by introducing the mini-superspace model of the bouncing universe with anisotropy in \ref{section_2}, whose classical dynamics has been studied in detail in \ref{classical_analysis}. Subsequently, we have explored the quantum aspects of our anisotropic bouncing model in \ref{WdW_section} and \ref{section_Lorentzian}, respectively. In particular, we have employed the approach based on the Wheeler-deWitt equation, in order to arrive at an analytical solution describing the wave function of the anisotropic bouncing universe in \ref{WdW_section}. While, in \ref{section_Lorentzian}, we verify our findings from solving the Wheeler-deWitt equation, using the Lorentzian path integral approach. Finally, in \ref{section_discussion}, we conclude with the summary of the results obtained and a discussion of future directions.

\emph{Notations and Conventions:} We have set the fundamental constant $c$ to unity and will follow the mostly positive signature convention, i.e., the metric of flat spacetime is taken to be $\textrm{diag.}(-,+,+,\cdots)$. 

\section{Bouncing model with shear: Mini-superspace approach}\label{section_2}

In this section we will provide the basic framework of the anisotropic bouncing model, which will be used extensively in the later parts of this work. Here, we consider the simplest class of anisotropic cosmological models, namely the Bianchi-I spacetimes. We shall parametrize the metric of the Bianchi-I spacetime as follows:
\begin{align}\label{def_Bianchi_I}
ds^{2}=-\frac{\mathcal{N}(t)^{2}}{q(t)^{4-3b}}dt^{2}
&+q(t)^{b}\Bigg[\exp\left(\sqrt{\frac{2\kappa}{3}}\left\{\theta_1(t)-\sqrt{3}\theta_2(t)\right\}\right)dx^{2}
\nonumber
\\
&\hskip 1.5 cm +\exp\left(\sqrt{\frac{2\kappa}{3}} \left\{\theta_1(t)+\sqrt{3}\theta_2(t)\right\}\right)dy^{2}
+\exp\left(-2 \sqrt{\frac{2\kappa}{3}}\theta_1(t)\right)dz^{2}\Bigg]~,
\end{align}
where, $b$ is a real number, which is arbitrary at this point and shall be fixed shortly to our convenience. It then follows, that $q(t)^{b/2}$ acquires the interpretation of the geometric mean of `scale factors' along the three independent spatial directions. As evident from the structure of the line element, due to the unknown functions $\theta_{1}(t)$ and $\theta_{2}(t)$, the spacetime is not isotropic. The advantage of the above parametrization is that the Einstein-Hilbert action, along with the Gibbons-Hawking-York boundary term\cite{York:1972sj,Gibbons:1976ue} on a $t=\textrm{constant}$ surface (for the corresponding scenario on a null surface, see \cite{Chakraborty:2019doh,Chakraborty:2016yna,Parattu:2016trq,Parattu:2015gga}), reduces to a simple form:
\begin{align}
\mathcal{S}_{\rm EH}&=\frac{1}{2\kappa}\int d^{4}x\sqrt{-g} R-\frac{1}{\kappa}\int_{\textrm{t=constant}}d^{3}y \sqrt{h}K
\nonumber
\\
&\hskip 2 cm =V_3\int dt\left[-\frac{M}{2\mathcal{N}}\dot{q}^2+\frac{q^2}{2\mathcal{N}}\left(\dot{\theta}^2_1+\dot{\theta}^2_2\right)\right]~,
\label{action_theta_12}
\end{align}
where, $M\equiv\{(3b^{2})/(2\kappa)\}$, with $\kappa=8\pi G$ ($G$ being the Newton's gravitational constant) and $V_3$ is the volume of the spatial slices. If, in addition, we assume periodic identifications along the spatial directions, i.e., along $x$, $y$ and $z$ coordinates, then the spatial slices will have the topology of a 3-torus and hence $V_{3}$ will correspond to the volume of the base torus of the spatial slices. 

Despite several possible avenues of realizing the bouncing scenario, we will be rather agnostic about the manner in which the specific bouncing we are considering is realized. Instead, to ensure that classical bouncing scenario is permitted, we shall merely introduce a suitable effective potential $U_{\rm eff}(q)$ to the above action. Finer details like, for instance, the origin of this potential from some effective field-theory description, shall not be our concern in this approach. Embracing this outlook, we find that the relevant action is given by
\begin{align}\label{action_ueff}
\mathcal{S}&=V_3\int dt\left[-\frac{M}{2\mathcal{N}}\dot{q}^2+\frac{q^2}{2\mathcal{N}}\left(\dot{\theta}^2_1+\dot{\theta}^2_2\right)+\mathcal{N}U_{\rm eff}(q)\right]~.
\end{align}
We emphasize, again, that the potential may be imagined as arising from either some modifications in the gravitational sector, consequent to, say, aspects of a theory of quantum gravity or, some `exotic' matter fields, or, both. However, as remarked before, we will simply study the consequences of the effective potential $U_{\rm eff}$ to quantum cosmology, without subscribing to any specific model for its microscopic origin. It may be convenient, in this picture, to imagine an effective perfect fluid density $\rho_{\rm eff}(q)$, whose presence introduce the same dynamics as that of the effective potential $U_{\rm eff}(q)$. One finds that this is achieved for $U_{\rm eff}=-\sqrt{-g}\rho_{\rm eff}(q)$ (for the metric described by \ref{def_Bianchi_I}, it follows that $\sqrt{-g}=\mathcal{N}(t)q^{3b-2}(t)$. In \cite{Rajeev:2021lqk}, two choices for the energy density of the perfect fluid were introduced, which lead to exactly solvable models of the bouncing universes. The corresponding effective potential $U_{\rm eff}(q)$, associated with these two models reduce to the following simple forms, respectively,
\begin{align}
U_{\rm eff}^{\rm (I)}(q)&=\rho_{0}(1-q);\qquad b\rightarrow b^{\rm (I)}=\frac{6}{6-n}~,
\label{intr_U_eff_I}
\\
U_{\rm eff}^{\rm (II)}(q)&=\rho_{0}(1-q^2); \quad b\rightarrow b^{\rm (II)}=\frac{8}{6-n}~,
\label{intr_U_eff_II}
\end{align}
where $\rho_0>0$, to ensure bouncing scenarios are allowed classically. On the right hand side of each of the equations above, we have displayed the corresponding choices of $b$, with $0<n<6$. For obvious reasons, the models described by the effective densities $U_{\rm eff}^{\rm (I)}$ and $U_{\rm eff}^{\rm (II)}$ were described as the `linear model' and the `quadratic model', respectively. It is worth examining the possible effective fluid description that can lead to the above potentials.  It turns out that the corresponding effective energy density takes the form $\rho^{(J)}_{\rm eff}(a)=\rho_{0}(a^{-n}-a^{-n_{J}})$, where $J=\{I,II\}$, with $n_{I}=(6+2n)/3$ and $n_{II}=(6+n)/2$. Consequently, when $a\ll 1$ the negative density component dominates while for $a\gg 1$ the effective density has the leading order behaviour $\rho^{(J)}_{\rm eff}\propto \rho_{0}a^{-n}$ (see \cite{Rajeev:2021lqk} for more details). Incidental, this desired $a\gg 1$ limit of the effective density manifests as an apparent divergent behaviour of $U_{\rm eff}(q)=-\sqrt{-g}\rho_{\rm eff}$ in the $q\gg 1$ limit. However, this divergent behaviour arises merely because the four-volume element $\sqrt{-g}$ scales as $q^{3b-2}$ and, hence, is perfectly innocuous. 

It will turn out that the quadratic model, with $M=(n-6)^{-2}(96/\kappa)$, is more analytically tractable in the presence of non-zero shear. Hence, we shall be focusing on the only the quadratic model, henceforth. This finishes the set up of the mini-superspace model for the bouncing scenario with shear. In what follows we will briefly touch upon the classical dynamics of the quadratic model, before taking up the quantum analysis of the bouncing scenario in the subsequent sections.

\section{Classical dynamics of the universe with shear}\label{classical_analysis}

In the previous section we have demonstrated the mini-superspace model of an anisotropic cosmological scenario, namely that of the Bianchi-I spacetime. This leads to the mini-superspace action, presented in \ref{action_ueff}. We will now describe the classical dynamics arising out of this action, with the effective potential given by \ref{intr_U_eff_II}. First of all, note that the action $\mathcal{S}$ in \ref{action_ueff} is invariant under translations in the $(\theta_{1},\theta_{2})$ plane, which implies the following integrals of motion:
\begin{align}\label{theta_momentum}
\Pi_{i}=\mathcal{N}^{-1}q^2\dot{\theta}_i=\textrm{conserved along classical solutions};\qquad i=1,2~.
\end{align} 
On the other hand, the constraint equation is obtained by varying the action $\mathcal{S}$ with respect to the lapse function $\mathcal{N}$, which yields:
\begin{align}\label{constraint_shear_old}
\frac{M}{2\mathcal{N}^2}\dot{q}^2+\rho_{0}\left(1-q^{2}\right)-\frac{\rho_{\theta}}{q^2}=0~,
\end{align}
where, we have defined the `shear density' $\rho_{\theta}\equiv(\Pi_{1}^2+\Pi_{2}^2)/2$. For later convenience, we shall henceforth work in a gauge in which $\dot{\mathcal{N}}=0$, i.e., $\mathcal{N}$ is independent of time. Then, for the special case of $\rho_{\theta}=0$, i.e., for zero shear, the classical solution for $q(t)$, or equivalently, for the scale factor $a(t)$, can be easily obtained by solving the constraint equation. This yields, 
\begin{align}
\label{a_II}
q(t)=\cosh\left(h_{n}\mathcal{N}t\right)~;\quad\quad a(t)=\left[\cosh\left(h_{n}\mathcal{N}t\right)\right]^{\frac{4}{6-n}}~,
\end{align} 
where we have traded off the constant $\rho_0$, appearing in the effective potential, for the parameter $h_{n}$, which is defined as:
\begin{align}\label{def_h}
h_{n}^{2}\equiv\frac{2\rho_{0}}{M}=\frac{\kappa}{48}(n-6)^{2}\rho_{0}~.
\end{align}
It is noteworthy that the classical solution for the scale factor $a(t)$ is appropriate for a symmetric, non-singular bouncing universe, since it never vanishes, rather it contracts, reaches a minimum and then keeps expanding. This behaviour of the scale factor clearly demonstrates how the effective density (or, equivalently, the effective potential $U_{\rm eff}$) gives rise to bouncing models of the universe.

\begin{figure}[h!]
\centering
\includegraphics[scale=0.37]{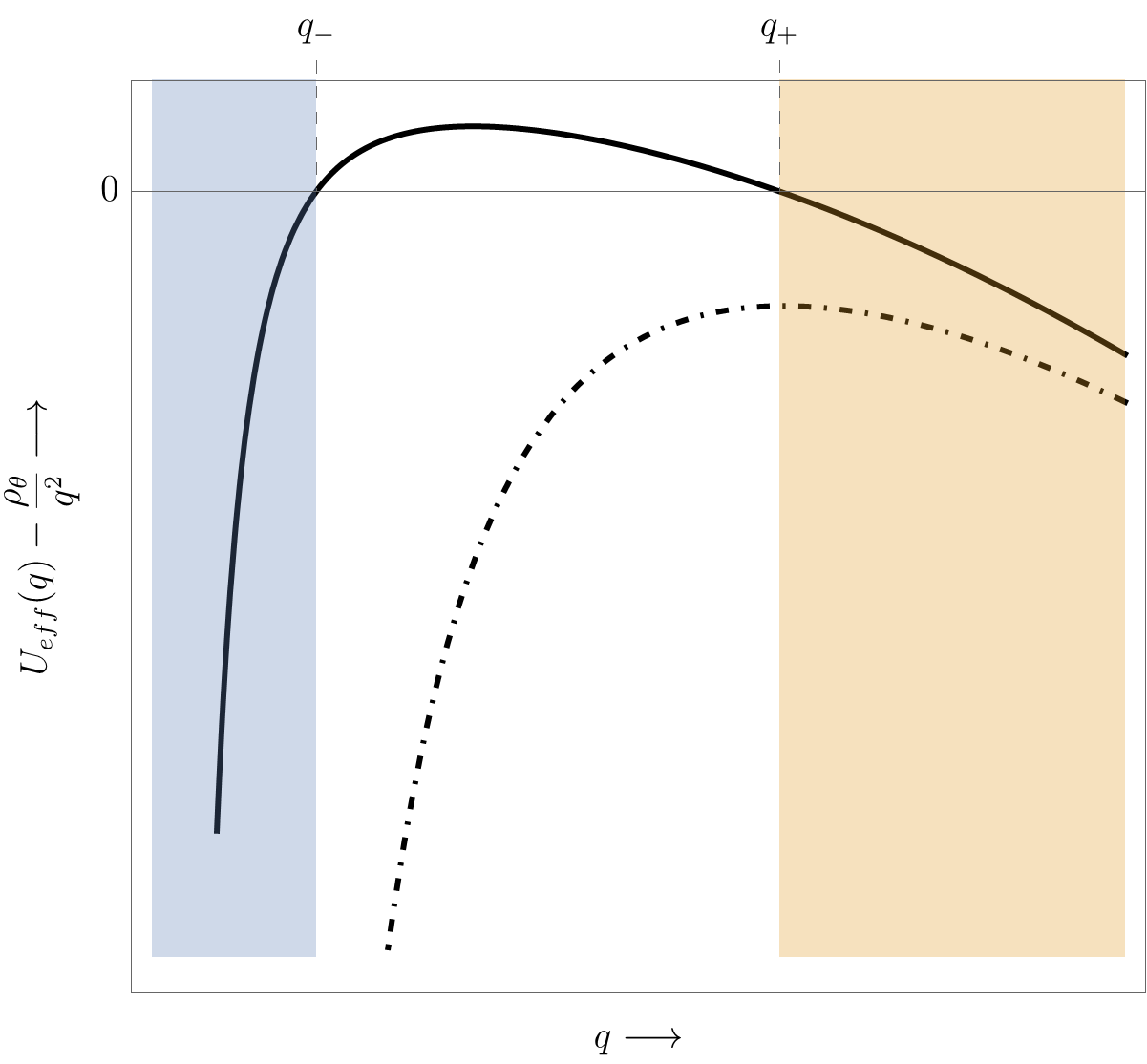}
\caption{The typical forms of the total effective potential $U_{\rm total}\equiv U_{\rm eff}-(\rho_{\theta}/q^{2})$ have been plotted, when (i) $0\leq(\rho_{\theta}/{\rho_0})<1/4$ (the thick curve) and (ii) $(\rho_{\theta}/\rho_{0})>1/4$ (the dot-dashed curve). The values $q_{-}$ and $q_{+}$, which are the classical turning points for $0\leq(\rho_{\theta}/{\rho_0})<1/4$ have also been depicted.}
\label{U_T_graph}
\end{figure}

In the presence of a non-zero $\rho_{\theta}$, there can be two distinct possible scenarios, depending on how large/small the value of $\rho_{\theta}$ is, relative to the density $\rho_{0}$. To see this clearly, let us rewrite \ref{constraint_shear_old} as, 
\begin{align}\label{constraint_shear}
\frac{M}{2\mathcal{N}^2}q^{2}\dot{q}^2=\rho_{0}\left(q^{2}-q_{+}^{2}\right)\left(q^{2}-q_{-}^{2}\right)~;\qquad q_{\pm}^{2}=\frac{1}{2}\pm \sqrt{\frac{1}{4}-\frac{\rho_{\theta}}{\rho_{0}}}~.
\end{align}
Thus for real $q_{\pm}$, since the left hand side of the above equation must be a positive definite quantity, we have two possibilities --- (a) $q>q_{+}$ and (b) $q<q_{-}$, i.e., classical turning points exist. As evident, for real values of $q_{\pm}$, the shear density must lie within the following range,
\begin{align}\label{bounce_allowed_range_II}
0\leq\frac{\rho_{\theta}}{\rho_0}< \frac{1}{4}~,
\end{align}  
and in this case the real classical turning points, at $q=q_{\pm}$, do exist. On the other hand, for $(\rho_{\theta}/\rho_{0})>(1/4)$, it follows that $q_{\pm}$ are complex and hence real classical turning points do not exist. Consequently, for the range of $\rho_{\theta}$, depicted in \ref{bounce_allowed_range_II}, a bouncing solution is indeed feasible classically. In particular, the solution $q(t)$ in the branch $q>q_{+}$, depicts a bouncing scenario. However, there can also be solutions in the other branch, satisfying $q<q_{-}$, for which classical solutions describing a big crunch (or, its time-reversed solution) also exists (see for example, \ref{U_T_graph}). In contrast, as emphasized above, when the shear density is such that $(\rho_{\theta}/\rho_{0})>1/4$, classical solutions describing smooth bouncing scenario cease to exist. 

\begin{figure}[h!]
\centering
\includegraphics[scale=0.3]{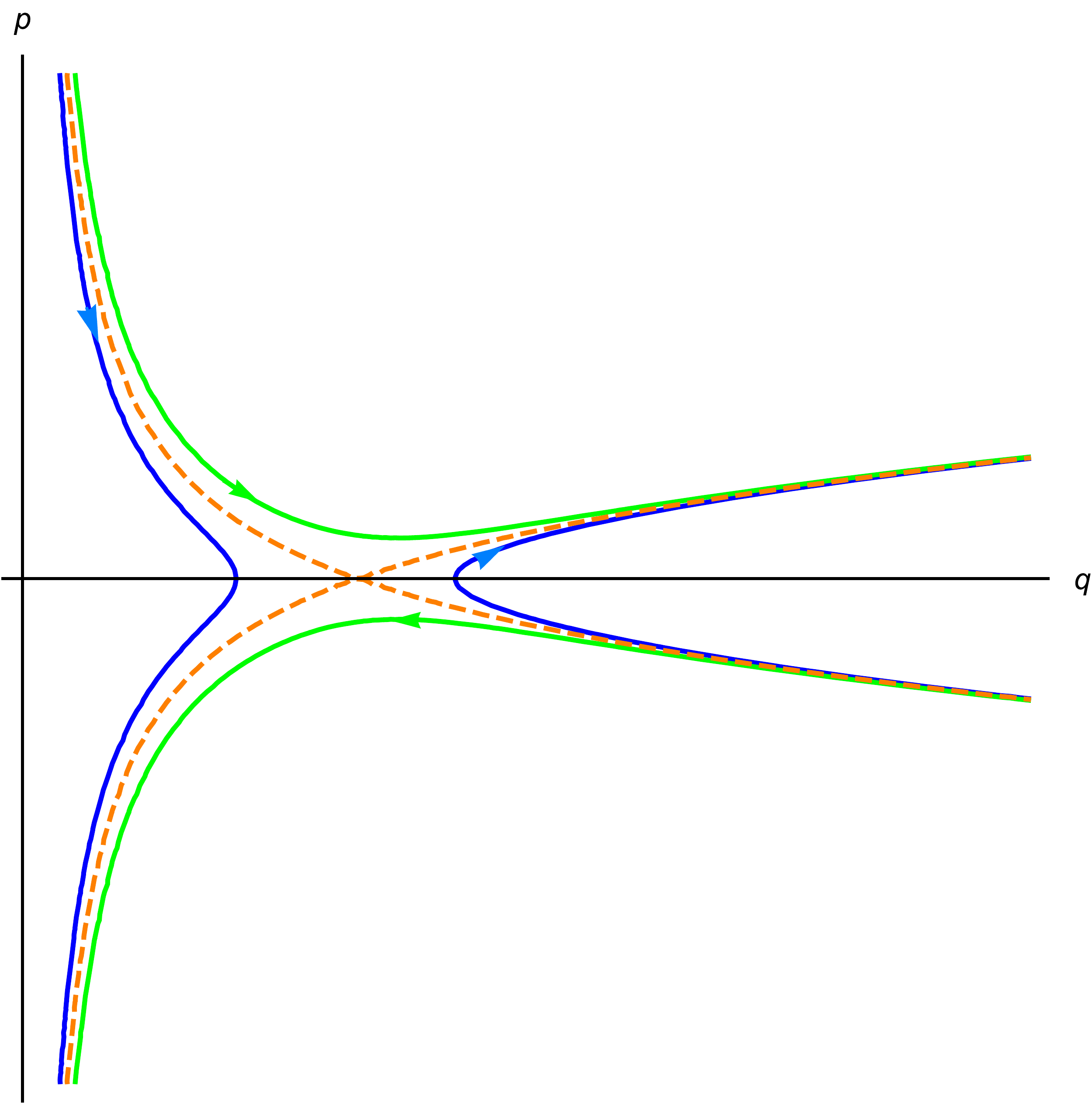}
\caption{The typical forms of classical solutions for two distinct ranges of $\rho_{\theta}$: blue curve represents the case in which $0<(\rho_{\theta}/\rho_0)<1/4$, while the green curves represents the case in which $(\rho_{\theta}/\rho_0)<1/4$. The arrows represent the direction of time. The dashed orange curves represent the separatrix (i.e., $(\rho_{\theta}/\rho_0)=1/4$).}
\label{phase_space_1}
\end{figure}

The two possibilities mentioned above are best illustrated by the phase space representation of the classical solutions. To this end, we start by rewriting the constraint equation, namely \ref{constraint_shear}, obtained by the variation of the mini-superspace action with respect to the Lapse function $\mathcal{N}$, in the phase space, as follows:
\begin{align}\label{Energy_function}
\mathcal{H}(p,q)\equiv\frac{p^2}{2M}+\rho_{0}\left(1-q^{2}\right)-\frac{\rho_{\theta}}{q^2}=0
\end{align}
where, $p\equiv \mathcal{N}^{-1}M\dot{q}$ is the \emph{negative} of the momentum conjugate to $q$. Hence, classical trajectories are represented by the $\mathcal{H}(p,q)=0$ contours in the $(p,q)$ plane, as we have illustrated in \ref{phase_space_1}. The blue contours correspond to the cases in which real classical turning points exist. The green contours, on the other hand, represent the cases in which there are no classical turning points. Notice that, in both cases, there are two disconnected parts for the $\mathcal{H}(p,q)=0$ contours. In case of the blue contours --- corresponding to the case in which $0<(\rho_{\theta}/\rho_0)< 1/4$---the two disconnected parts have the following interpretations: (i) the left part corresponds to a universe that expands from singularity to a maximum size and then starts contracting back to the singularity; we shall refer to this scenario as the expanding-contracting universe and (ii) the right part describes a bouncing universe. Therefore, when $0<(\rho_{\theta}/\rho_0)< 1/4$, depending on the initial conditions, a classical solution may either correspond to an expanding-contracting universe or a bouncing universe, and any smooth transition from the former to the latter or vice-versa is forbidden classically. In case of the green contours, corresponding to the case in which $(\rho_{\theta}/\rho_0)> 1/4$ --- the two disconnected parts represent the following scenarios: (i) the top part corresponds to an ever-expanding universe and (ii) the bottom part describes an ever-contracting universe. Hence, when $(\rho_{\theta}/\rho_0)> 1/4$, depending on the initial conditions, the classical solutions correspond to either an ever-expanding or an ever-contracting universe, and any smooth transition from the former to the latter or vice-versa is forbidden. However, we shall see shortly that quantum mechanically transitions between disconnected classical solutions are allowed in the form of tunnelling and over-the-barrier reflection. 

So far, we have presented the nature of the classical solutions in the presence of shear. For completeness we present below the exact solutions for $q(t)$, for any value of the shear density $\rho_{\theta}$, by solving \ref{constraint_shear}. For that purpose, one may introduce a change of variable, namely $x(t)\equiv q(t)^{2}$, in terms of which \ref{constraint_shear} can be re-expressed as,
\begin{align}
\frac{M}{8\mathcal{N}^2}\dot{x}^2=\rho_{0}\left(x-x_{+}\right)\left(x-x_{-}\right)~,
\end{align}
where, $x_{\pm}\equiv q_{\pm}^{2}$. The above equation can be immediately solved in terms of the Hyperbolic functions, when the shear density is within the range $0<(\rho_{\theta}/\rho_{0})<(1/4)$ (this corresponds to the blue curves of \ref{phase_space_1}), for which the solution for $q(t)$ takes the following form, 
\begin{align}\label{solutions_b}
q_{\rm B}(t)= \frac{1}{\sqrt{2}}\sqrt{1+\cos(2\gamma)\cosh\left(2h_n\mathcal{N}t\right)}~,
\\
\label{solutions_ec}
q_{\rm EC}(t)= \frac{1}{\sqrt{2}}\sqrt{1-\cos(2\gamma)\cosh\left(2h_n\mathcal{N}t\right)}~,
\end{align}
where, the subscript `${\rm B}$' denotes the bouncing solution (blue curve on the right in \ref{phase_space_1}) and `${\rm EC}$' denotes the expanding-contracting phase (blue curve on the left in \ref{phase_space_1}), and we have also defined a parameter $\gamma$ such that $\rho_{\theta}\equiv(\rho_0/4)\sin^2(2\gamma)$ with $\gamma\in[0,\frac{\pi}{4}]$. Similarly, the solutions corresponding to the green curves of \ref{phase_space_1} (i.e., when $(\rho_{\theta}/\rho_0)>1/4$) are found to be: 
\begin{align}
q_{\rm E}(t)=\frac{1}{\sqrt{2}}\sqrt{1+\sinh\left(2 h_n\mathcal{N} t\right)\sinh(2\delta)}~,
\\
q_{\rm C}(t)=\frac{1}{\sqrt{2}}\sqrt{1-\sinh\left(2 h_n\mathcal{N} t\right)\sinh(2\delta)}~,
\end{align}
where, the subscript ${\rm E/C}$ denotes the green curves on the top/bottom of \ref{phase_space_1}, corresponding to `ever-expanding' and `ever-contracting' universes, respectively. We have also introduced the parameter $\delta$ through the definition $(\rho_{\theta}/\rho_0)\equiv(1/4)\cosh^2(2\delta)$ with $\delta>0$. 

This finishes the discussion regarding classical solutions in the presence of shear. As we have seen, classically, bouncing models are allowed only for small shear density, i.e., if the shear density $\rho_{\theta}$ satisfies the condition $(\rho_{\theta}/\rho_{0})\leq (1/4)$. Larger values of shear, classically prohibits the existence of bouncing scenarios. In what follows, we will show that even when the shear density is large, it is possible to have a bouncing scenario, when quantum effects are taken into account. This result will be achieved by solving the Wheeler-deWitt equation and also using path-integral techniques. 

\section{Wheeler-deWitt equation and its solution with shear}\label{WdW_section}

In this section, we will study the quantum aspects of anisotropic cosmological models with a certain perfect fluid, such that in the absence of anisotropy a bouncing solution is realized, classically. As we have seen, the presence of shear has a very adverse effect on the bouncing models. In particular, larger values of the shear density forbids a bouncing scenario. We would like to analyze this problem from the inclusion of quantum effects and the first direction to achieve the same is to solve the Wheeler-deWitt equation. This is what we perform in this section. 

The Wheeler-deWitt equation is obtained by replacing the momenta in the Hamiltonian constraint by the corresponding operators. For example, in this context, $p=-i\hbar(\partial/\partial q)$ and $\Pi_{i}=-i\hbar(\partial/\partial \theta_{i})$, such that for the system of current interest, described by the action in \ref{action_theta_12}, we obtain the Wheeler-deWitt equation to take the following form:
\begin{align}\label{WdW_1}
\left[-\frac{\hbar^{2}}{2MV_{3}}\partial^{2}_{q}+\frac{\hbar^{2}}{2V_{3}q^{2}}\left(\partial^{2}_{\theta_{1}}+\partial^{2}_{\theta_{2}}\right)+V_{3}\rho_{0}\left(1-q^{2}\right)\right]\Psi(q,\theta_{1},\theta_{2})=0~,
\end{align} 
which we will solve by an appropriate redefinition of the variables in what follows. As an aside, we would like to point out an interesting way of rewriting the above equation. This is achieved by introducing the following notation for the coordinates of the mini-superspace: $(q^{0},q^{1},q^{2})=(q,\theta_{1},\theta_2)$ and introducing the mini-superspace metric as,
\begin{align}\label{metric_mss}
ds^{2}_{\rm mSS}=G_{AB}dq^{A}dq^{B}\equiv-MV_{3}dq^{2}+V_{3}q^{2}\left(d\theta_{1}^{2}+d\theta_{2}^{2}\right)~,
\end{align}  
where the indices $(A,B,\cdots)$ run over $(0,1,2)$, respectively. In terms of the above mini-superspace metric $G_{AB}$, the Wheeler-deWitt equation can be written compactly as:
\begin{align}
\left[\frac{\hbar^2}{2\sqrt{-G}}\partial_{q^A}\left(\sqrt{-G}G^{A B}\partial_{q^B}\right)+V_{3}\rho_{0}\left(1-q^{2}\right)\right]\left(\frac{\Psi(q,\theta_1,\theta_2)}{q}\right)=0~.
\end{align}
As evident, the Wheeler-deWitt equation presented above, is very much like the Klein-Gordon equation with a negative mass squared term in a spacetime, whose metric is given by \ref{metric_mss}. Incidentally, the mini-superspace metric $G_{AB}$, introduced in \ref{metric_mss}, describes an FLRW universe sourced by a perfect fluid with equation of state $p=-(\rho/3)$ and $q$ playing the role of time. For this work, we will content ourselves by merely pointing out this curious observation, while it may be worthwhile to explore if there are any more utility to the same, which we hope to report in a separate future work.

Returning to our primary task, i.e., solving the Wheeler-deWitt equation, as a first step we may use the fact that the potential does not depend on $\theta_{1}$ and $\theta_{2}$. This in turn suggests, as in the classical scenario, translational symmetry in the $(\theta_{1},\theta_{2})$ plane, thus enabling us to seek separable solutions of the following form:
\begin{align}\label{separate_variable}
\Psi_{\vec{\Pi}}(q,\vec{\theta})=\psi_{\vec{\Pi}}(q)e^{i\frac{V_3}{\hbar}\vec{\Pi}.\vec{\theta}}~,
\end{align}  
where, we have introduced the notation $\vec{\Pi}\equiv(\Pi_1,\Pi_2)$ and $\vec{\theta}\equiv(\theta_1,\theta_2)$. Then, the wave function $\psi_{\vec{\Pi}}(q)$ satisfies the following differential equation:
\begin{align}\label{WdW_2}
\left[-\frac{\hbar^2}{2M V^2_3}\partial^2_{q}+U_{\rm total}(q)\right]\psi_{\vec{\Pi}}(q)=0~,
\end{align} 
where, the `total potential' $U_{\rm total}(q)$ takes the following form,
\begin{align}
U_{\rm total}(q)=\rho_{0}\left(1-q^{2}\right)-\frac{|\vec{\Pi}|^2}{2q^2}~,
\end{align} 
with the following shorthand notation, $|\vec{\Pi}|^{2}\equiv \Pi_{1}^{2}+\Pi_{2}^{2}$. From a direct comparison with our analysis of classical dynamics, for instance, with \ref{Energy_function}, we can immediately see that the shear density corresponds to $\rho_{\theta}=|\vec{\Pi}|^2/2$. Thus the Wheeler-deWitt equation has been reduced to a single ordinary differential equation in the variable $q$. In what follows we will solve this differential equation in order to derive the wave function of the universe associated with a potential as fit for the bouncing models, but in the presence of shear. However, before going into the details of the solution, we will present the basic physical premise to expect out of this analysis. 

We have already noticed in \ref{classical_analysis}, that a bouncing scenario is classically feasible only up to a maximum value of $\rho_{\theta}$ (in particular, for $0<(\rho_{\theta}/\rho_{0})<(1/4)$). When the shear density exceeds this critical value of $(\rho_{0}/4)$, a contracting phase inevitably ends in a singularity. Conversely, when the shear density is small enough, a contracting phase, with large enough value of the scale factor ($q>q_{+}$, see \ref{constraint_shear} for details), is always succeeded by a bounce. However, taking a cue from \cite{Rajeev:2021lqk}, we may expect interesting new scenarios to arise in the quantum cosmological treatment. In particular, we expect these two key features, which are purely quantum effects, to arise in the present scenario:
\begin{enumerate}

\item There is a finite probability for a contracting universe, with large enough value of the scale factor, to eventually collapse to singularity, \textit{even when} the shear $\rho_{\theta}$ is small enough for a classical bounce to be feasible.

\item There is a finite probability for a contracting phase to be followed by an expanding phase, i.e., bounce to happen, \textit{even when} the shear $\rho_{\theta}$ is large enough to forbid a classical bouncing scenario.

\end{enumerate}
One can imagine these effects as, respectively, the consequences of `tunnelling' and `over-the-barrier-reflection' that are familiar in quantum mechanics. The exact solutions of \ref{WdW_2} must be able to describe these phenomena, by providing analytical expressions for the above probabilities. Let us briefly examine the requisite characteristics of the wavefunction $\Psi_{\vec{\Pi}}(q)$, which will be appropriate for our analysis. Motivated by the standard scattering problem in quantum mechanics, we demand that $\Psi_{\vec{\Pi}}(q)$ has following asymptotic behaviour:
\begin{align}\label{asym_behaviour_gen}
\Psi\sim\begin{cases}
\Psi_{\rm contract}+A_{\rm bounce}\,\,\Psi_{\rm expand}&;\qquad \textrm{scale factor}\gg 1\\
A_{\rm collapse}\Psi_{\rm collapse}&;\qquad \textrm{scale factor}\ll 1
\end{cases}
\end{align}
where, $\Psi_{\rm contract/expand/collapse}$ corresponds to a wave function describing contracting/expanding/collapsing phase of the universe. The coefficients $A_{\rm bounce}$ and $A_{\rm collapse}$ are, respectively, the amplitudes for bounce and collapse to occur. For $(\rho_{\theta}/\rho_{0})<(1/4)$, it follows that classically, $A_{\rm collapse}$ should be zero and $A_{\rm bounce}$ should be unity. However, quantum mechanically we expect $A_{\rm collapse}$ to be different from zero and $A_{\rm bounce}$ to be different from unity. Similarly, for $(\rho_{\theta}/\rho_{0})>(1/4)$, it follows that classically, $A_{\rm collapse}$ should be unity and $A_{\rm bounce}$ should be zero. Again, quantum mechanically, we expect both $A_{\rm collapse}$ and $A_{\rm bounce}$ to inherit values different from those suggested by the classical computation. 

We have seen that the absolute value of $A_{\rm bounce}$ can, in general, be different from unity, when quantum mechanical effects are taken into account. Note that, when viewed in light of the aforementioned settings, the original no-boundary state of Hartle and Hawking can also be interpreted as one in which $|A_{\rm bounce}|$ is unity. The same also holds true for the analogue of the no-boundary wave function for the isotropic bouncing models discussed in \cite{Rajeev:2021lqk}. Therefore, the departure of $A_{\rm bounce}$ from unity, introduced by anisotropy, might lead to interesting deviations from the no-boundary state features. This, in turn, might have important consequences for the present day cosmology. Hence, we shall now analyse the wave function of an anisotropic bouncing universe in greater details.

\subsection{An exact wave function of the universe with shear}\label{section_WdW_solution}

Having described above the key physics inputs, in this section we will derive an exact wave function of the universe, in the presence of shear, by solving \ref{WdW_2}. For this purpose and also for future manipulations, it is convenient to rewrite \ref{WdW_2} in the following form:
\begin{align}\label{WdW_II}
\left[-\frac{\partial^{2}}{\partial q^{2}}+\alpha_{n}^{2}\left(1-q^{2}-\frac{\sigma^2}{4q^2}\right)\right]\psi_{\vec{\Pi}}(q)&=0~,
\end{align} 
where, we have introduced the parameters $\sigma$ and $\alpha_{n}$ through the following definitions:
\begin{align}\label{sigma_def}
\frac{1}{4}\sigma^{2}&\equiv\frac{\rho_{\theta}}{\rho_0}\\
\label{alpha_def_1}
\alpha^2_{n}&\equiv\frac{2MV_{3}^{2}\rho_{0}}{\hbar^{2}}=\frac{M^{2}V_{3}^{2}h_{n}^{2}}{\hbar^{2}}=\frac{16}{9}\left(\frac{n}{6}-1\right)^{-2}\left(\frac{\rho_0}{\rho_{\rm Planck}}\right)\left(\frac{V_3^2}{V_{\rm Planck}^2}\right)
\end{align}
where, we have defined the `Planck length', `Planck density' and `Planck volume', respectively, as $\ell_{\rm Planck}=\sqrt{\hbar\kappa/3}$, $\rho_{\rm Planck}\equiv \hbar\ell_{\rm Planck}^{-4}$ and $V_{\rm Planck}=\ell_{\rm Planck}^3$. The reduced Wheeler-deWitt equation, as presented in \ref{WdW_II}, resembles the Schr\"{o}dinger equation with a harmonic oscillator and an inverse harmonic oscillator potential, which with appropriate substitution can be exactly solved (for more details, see \ref{App_Alt}. This is what we will do in the subsequent discussion. After deriving the wave function, we will apply the appropriate boundary conditions, presented in \ref{asym_behaviour_gen}, in order to determine the probabilities of bounce and collapse in the presence of shear. 

The first step in solving the reduced Wheeler-deWitt equation is to make the following change of variable, $z=i\alpha_{n}q^{2}$ and also the following redefinition of the wave function, $\psi_{\vec{\Pi}}(q)\equiv z^{-\frac{1}{4}}\bar{\psi}(z)$. In terms of this newly defined wave function $\bar{\psi}(z)$, the reduced Wheeler-deWitt equation, presented in \ref{WdW_II}, can be rewritten as:
\begin{align}\label{Conf_Hyp_shear}
\frac{d^2\bar{\psi}}{dz^2}+\left(-\frac{1}{4}+\frac{\mu}{z}+\frac{\frac{1}{4}-\nu^2}{z^2}\right)\bar{\psi}=0~,
\end{align}
where, the constants $\mu_{1}$ and $\mu_{2}$, appearing in the above differential equation, have the following expressions,
\begin{align}
\mu&=\frac{i\alpha_n}{4}
\label{def_mu1}
\\
\nu&=\begin{cases}
\frac{i}{4}\sqrt{\alpha_n^2\sigma^2-1}\quad&;\qquad\alpha_n\sigma\geq 1\\
\frac{1}{4}\sqrt{1-\alpha_n^2\sigma^2}\quad&;\qquad\alpha_n\sigma< 1
\end{cases}
\label{def_mu2}
\end{align}
where, $\sigma$ and $\alpha_{n}$ are defined in \ref{sigma_def} and \ref{alpha_def_1}, respectively. We seek for a solution of \ref{Conf_Hyp_shear} that has the asymptotic behaviour described in \ref{asym_behaviour_gen}. As we will explicitly demonstrate, the desired solution with the correct asymptotic behaviour is the Whittaker's function, which can be expressed as,
\begin{align}\label{sol_psi_bar}
\bar{\psi}(z)=M_{\mu,\nu}(z)~.
\end{align} 
In order to verify that this is, in fact, the desired solution, let us look at its asymptotics. Note that, large scale factor implies large $q$ and vice versa. Thus we need to consider $q\rightarrow 0$ and $q\rightarrow \infty$ limit of the Whittaker's function, $M_{\mu,\nu}(z)$, keeping in mind that $z=i\alpha_{n}q^{2}$. Such an asymptotic expansion yields,
\begin{align}
\psi_{\vec{\Pi}}(q)\approx
\begin{cases}
\left(i\alpha_nq^2\right)^{\nu+\frac{1}{4}}\quad&;\quad q\rightarrow 0
\\
\mathcal{A}_{+}\left(i\alpha_nq^2\right)^{-\frac{1}{4}}e^{\frac{i}{2}\alpha_n\left(q^2-\log q\right)}
+\mathcal{A}_{-}\left(i\alpha_nq^2\right)^{-\frac{1}{4}}e^{-\frac{i}{2}\alpha_n\left(q^2-\log q\right)} \quad&;\quad q\rightarrow\infty
\label{largeq_limit}
\end{cases}
\end{align}
where, the constants $\mathcal{A}_{+}$ and $\mathcal{A}_{-}$ have the following expressions in terms of the functions $\mu_{1}$ and $\mu_{2}$, defined in \ref{def_mu1} and \ref{def_mu2}, respectively,
\begin{align}\label{aplusminus}
\mathcal{A}_{+}=\frac{e^{\frac{\pi}{2}|\mu|}\left(\alpha_n\right)^{-i|\mu|}\Gamma(1+2\nu)}{\Gamma(1/2+\nu-i|\mu|)} \quad;
&&
\mathcal{A}_{-}=\frac{e^{\frac{\pi}{2}|\mu|}\left(\alpha_n\right)^{i|\mu|}e^{i\pi\left(\nu+\frac{1}{2}\right)}\Gamma(1+2\nu)}{\Gamma(1/2+\nu+i|\mu|)}~.
\end{align}
In order to interpret the asymptotic form of the solutions, it will be instructive to look at the Jacobi form of the action. Since the quantum cosmology effectively boils down to a single particle quantum mechanics problem with an effective potential $U_{\rm total}(q)$, it follows that in the present context asymptotically the Jacobi action reads,
\begin{align}
\frac{S_{\pm}(q)}{\hbar}&=\pm\frac{1}{\hbar}\int\sqrt{2M |U_{\rm total}(q)|}dq
\nonumber
\\
&\approx \pm\frac{\alpha_n}{2}\left(q^2-\log q\right)\quad;\qquad q\rightarrow\infty~.
\end{align}
Therefore, in the large $q$ limit, the wave function $\psi_{\vec{\Pi}}(q)$ with shear, given in \ref{largeq_limit}, can also be re-expressed as:
\begin{align}\label{WKB}
\psi_{\vec{\Pi}}(q)\propto \mathcal{A}_{+}\frac{e^{\frac{i}{\hbar}S_{\rm+}}}{\sqrt{|\partial_{q}S_{\rm+}|}}+\mathcal{A}_{-}\frac{e^{\frac{i}{\hbar}S_{\rm-}}}{\sqrt{|\partial_{q}S_{\rm-}|}}~.
\end{align}
The Jacobi actions, namely $S_{\rm\pm}$(q), appearing in the each of the independent exponential terms in \ref{WKB} can be, respectively, attributed to the classical solution defined by $-\partial_{q}S_{\rm\pm}=MV_3\dot{q}$. A positive/negative value for $\dot{q}$ clearly corresponds to an expanding/contracting universe. This implies that $S_{\rm +}$ describes a contracting solution, while $S_{\rm -}$ describes an expanding solution. In light of this, a comparison with \ref{asym_behaviour_gen} and \ref{WKB} shows that $\psi_{\vec{\Pi}}(q)$ has the desired form in the limit of large scale factor (equivalently, $q\gg 1$ limit). In a similar manner, one can also verify that $\psi_{\vec{\Pi}}(q)$ also has the desired form in the limit of small values of scale factor (equivalently, $q\ll 1$ limit).\footnote{For examining the $q\rightarrow0$ limit, it is more convenient to work with the variable $x=\log q$ and the wave function $\phi_{\Pi}(x)\equiv e^{-x/2}\psi_{\vec{\Pi}}(e^x)$. Then, the particular solution to the corresponding Wheeler-deWitt equation, that satisfies our boundary condition near $q\rightarrow 0$, takes the simple form $\phi_{\Pi}(x)\approx {\rm (constant)} e^{2\mu x}$, in the limit $x\rightarrow-\infty$.}.

\begin{figure}[h]
\centering
\includegraphics[scale=0.8]{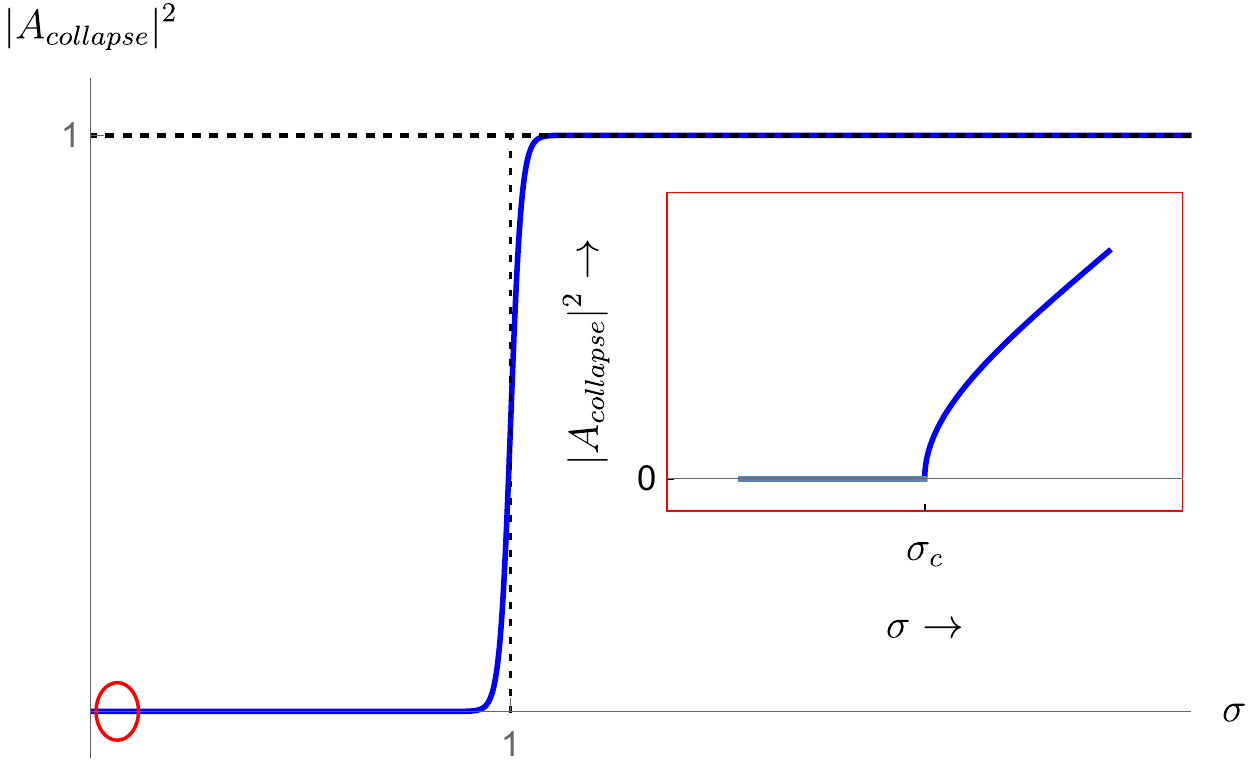}
\caption{The probability for collapse $\mathcal{T}(\sigma)$ (or, equivalently $|A_{\rm collapse}|^{2}$) has been plotted as a function of $\sigma$, when $\alpha_{n}\gg 1$. It shows that initially the probability stays at zero, but after certain critical value it increases drastically. The region inside the red circle is enlarged and plotted in the inset, showing the behaviour of $\mathcal{T}(\sigma)$ near the critical value $\sigma_c=4\rho_c/\rho_0$.}
\label{Exact_T}
\end{figure}

In particular, while the term with $\mathcal{A}_{+}$ as the coefficient corresponds to the classical contracting solution, the term with $\mathcal{A}_{-}$ as the coefficient corresponds to the classical expanding solution. Thus, the amplitude to bounce $A_{\rm bounce}$, as defined in \ref{asym_behaviour_gen}, can be easily read off as the ratio of $\mathcal{A}_{-}$ and $\mathcal{A}_{+}$:
\begin{align}
A_{\rm bounce}=\frac{\mathcal{A}_{-}}{\mathcal{A}_{+}}
=\left(\alpha_n\right)^{2i|\mu|}e^{i\pi\left(\nu+\frac{1}{2}\right)}\frac{\Gamma(1/2+\nu-i|\mu|)}{\Gamma(1/2+\nu+i|\mu|)}~.
\end{align} 
Substituting the expressions for $\mu_{1}$ and $\mu_{2}$ in terms of $\alpha_{n}$ and $\sigma$, the probability to bounce, which we shall denote by $\mathcal{R}$, then evaluates to:
\begin{align}\label{exact_R}
\mathcal{R}(\sigma)\equiv|A_{\rm bounce}|^2=\begin{cases}
\frac{\exp\left[-\frac{\pi}{2}\left(\sqrt{\alpha_{n}^{2}\sigma^{2}-1}+\alpha_{n}\right)\right]+1}{\exp\left[\frac{\pi}{2}\left(\sqrt{\alpha_{n}^{2}\sigma^{2}-1}-
\alpha_{n}\right)\right]+1}\quad&;\qquad \alpha_{n}\sigma\geq 1
\\
1\quad&;\qquad \alpha_{n}\sigma<1
\end{cases}
\end{align}
Then, the probability for collapsing to singularity, which we shall denote by $\mathcal{T}$, can be readily computed via the following relation:
\begin{align}\label{exact_T}
\mathcal{T}(\sigma)\equiv|A_{\rm collapse}|^2=1-\mathcal{R}(\sigma)~,
\end{align}
where, $\mathcal{R}(\sigma)$ is the probability to bounce, whose expression is given in \ref{exact_R}. The behaviour of $\mathcal{T}$, or, equivalently of $|A_{\rm collapse}|^{2}$ as a function of $\sigma$ is shown in \ref{Exact_T}, from which the behaviour of $\mathcal{R}$ can also be inferred, since it is simply $(1-\mathcal{T})$, as \ref{exact_T} suggests. 

The expression for $\mathcal{R}$ from \ref{exact_R}, suggests the following intriguing possibility: $\mathcal{R}$ is \textit{exactly} equal to unity (equivalently, $\mathcal{T}$ is \textit{exactly} zero) for $\sigma$ less than a critical value $\sigma_c$, satisfying $\alpha_{n}\sigma_{c}=1$. This, in turn, means that there is a critical value $\rho_{\theta,c}$ of the shear density above which the quantum cosmological effect starts to reveal itself. It follows, from the definition of $\sigma$ and $\alpha_{n}$, given by \ref{alpha_def_1}, that
\begin{align}
\frac{\rho_{\theta,c}}{\rho_{\rm Planck}}=\frac{9}{64}\left(\frac{n}{6}-1\right)^2\left(\frac{V_{\rm Planck}^{2}}{V_{3}^{2}}\right)~.
\end{align} 
Note that for small values of $\sigma$, the parameter $V_{3}$ is the volume and $\rho_{\theta}$ is the shear density, both evaluated \textit{at} the bounce. Moreover, for a bouncing universe, the volume and shear, respectively, acquire the minimum and maximum values at the bounce. Therefore, physically, the parameters $V_3$ and $\rho_{\theta}$ is a measure of maximum value of volume and minimum value of shear contribution to density, respectively. Hence, one can say that the quantum cosmological effects become pronounced when the maximum value of the classical shear density $\rho_{\theta}$ exceeds the critical value $\rho_{\theta,c}$. Since the critical shear density $\rho_{\theta,c}$ is a significant quantity in the context of bouncing cosmology with shear, it is worth highlighting the behaviour of  the probability of collapse, i.e., $\mathcal{T}$ near $\rho_{\theta,c}$, which takes the following form,
\begin{align}\label{T_exact}
\mathcal{T}(\sigma\approx\sigma_c)\approx\begin{cases}
0\qquad&;
\qquad \rho_{\theta}<\rho_{\theta,c}
\\
\frac{\pi }{\exp\left[\frac{\pi \alpha_{n}}{2}\right]+1}\left(\frac{\rho_{\theta}}{\rho_{\theta,c}}-1\right)^{\frac{1}{2}}\qquad&;
\qquad\rho_{\theta}>\rho_{\theta,c}~.
\end{cases}
\end{align}  
The behaviour of $\mathcal{T}(\sigma)$, the probability of collapse, near the critical value of shear $\rho_{\theta,c}$, is visualised in the inset of \ref{Exact_T}. In the light of \ref{T_exact} and also from \ref{Exact_T}, it follows that the form of $\mathcal{T}(\sigma)$ is as though there is a phase-transition (from `classical-to-quantum') happening at $\rho_{\theta}=\rho_{\theta,c}$; while the probability of collapse is \textit{exactly} zero for $\rho_{\theta}<\rho_{\theta,c}$, like in the classical case, the same increases \textit{sharply} as we increase $\rho_{\theta}$ slightly from its critical value $\rho_{\theta,c}$. 

It is reasonable to believe that phase-transition like feature is universal to a large class of anisotropic bouncing scenarios. This is because the key ingredient for this peculiar behaviour of $\mathcal{T}(\sigma)$ seems to be the presence of the $(1/q^2)$ term in the potential that dominates near $q\rightarrow0$ and the presence of shear is guaranteed to provide such a term provided the rest of the potential is suitable well behaved near the singularity. It is well know that a particle in $(g/q^2)$ potential has a scaling symmetry and upon quantization this symmetry gets broken at a critical value of the coupling $g=g_c$ \cite{Gupta:1993id}. The phase-transition like behaviour of $\mathcal{T}(\sigma)$, as described by \ref{T_exact}, is nothing but a manifestation of this quantum anomalous behaviour one expects in the presence of $(g/q^2)$ potential. It is worth mentioning that the relevance of conformal symmetry in the mini-superspace has been appreciated in literature before (for instance, see \cite{BenAchour:2019ywl}). However, it would be worthwhile to investigate the deeper significance of a scaling symmetry and its breaking in the context of quantum cosmology in finer details, which we shall reserve for an upcoming paper.  

Another limit, corresponding to $\alpha_{n}\gg 1$, is worth examining in the present context. This can actually be considered as equivalent to the semi-classical limit, since from \ref{alpha_def_1} it is clear that one can realize the $\alpha_n\gg1$ limit by fixing $\rho_{0}V_{3}^2$ to be much larger than the corresponding Planck quantity $\rho_{\rm Planck}V_{\rm Planck}^2$. From \ref{exact_R} it follows that the probability of collapse and bounce, in the $\alpha_n\gg 1$ limit, take the following simple forms, respectively:
\begin{align}\label{T_approx}
\mathcal{T}(\sigma)\approx\frac{e^{\frac{\pi \alpha_n}{2}(\sigma -1)}}{e^{\frac{\pi  \alpha_n}{2}(\sigma -1)}+1}, &&\mathcal{R}(\sigma)\approx\frac{1}{e^{\frac{\pi  \alpha_n}{2}(\sigma -1)}+1}~ \quad;\quad \frac{1}{\alpha_n}<\sigma<\infty\textrm{ and }\alpha_n\gg 1.
\end{align}
As evident, for $\alpha_{n}\gg 1$, the probability of bounce $\mathcal{R}$ is exponentially small, though non-zero, when $\sigma>1$ while the probability of collapse is very close to unity. On the other hand, in the same semi-classical limit $\alpha_{n}\gg 1$, the probability of collapse $\mathcal{T}$ is exponentially small, though non-zero, when $\sigma<1$ while the probability of bounce is very close to unity. We will now provide interpretation for the above expressions of $\mathcal{R}$ and $\mathcal{T}$ in terms of quantum tunnelling and over-the-barrier reflection, thereby, also associating the physical meaning of the wave function $\Psi_{\vec{\Pi}}(q)$. 

\subsection{Semi-classical approximation and the interpretation of the wave function}\label{WdW_semiclassical}

In this section, we will provide an alternative interpretation of the wave function $\psi_{\vec{\Pi}}(q)$ as tunnelling from one classically allowed region to another. As we have already demonstrated in \ref{classical_analysis}, the classical solutions are represented by disconnected curves in the $p-q$ plane (also see \ref{phase_space_1}). Given that our main interest is in the bouncing scenario, we will mainly consider the case in which the shear density $\rho_{\theta}$ falls within the range $0\leq(\rho_{\theta}/\rho_{0})<(1/4)$. In this case, as we have seen in \ref{classical_analysis}, there are two possibilities --- (a) the classical solution describes a bouncing universe, if it started contracting from a sufficiently large size, (b) the classical solution describes a big crunch scenario, as the universe expands from singularity, reaches a maximum size and then contracts back to the singularity. In what follows, we will show how to connect these disjoint classical regimes together, which in turn will provide an interpretation for the wave function in terms of tunnelling. 

As evident from \ref{solutions_b} and \ref{solutions_ec}, the bouncing solution $q_{\rm bounce}(t)$ describes a universe that expands to arbitrarily large values of the scale factor, but has a lower bound for the size corresponding to $q_{\rm bounce}(t)= q_{+}$. In contrast, the expanding-contracting solution $q_{\rm EC}(t)$ has no lower bound to its size, since it can reduce all the way to zero, but acquires an upper bound to the size corresponding to $q_{\rm EC}(t)=q_{-}$. Since $q_{+}^{2}+q_{-}^{2}=1$, it is convenient to parametrize $q_{+}$ and $q_{-}$, as $\cos \gamma$ and $\sin \gamma$, respectively, where $\gamma\in[0,\frac{\pi}{4}]$. It follows, that $\gamma=0$ corresponds to $(\rho_{\theta}/\rho_{0})=0$, while $\gamma=(\pi/4)$ yields $(\rho_{\theta}/\rho_{0})=(1/4)$. To summarize, the ranges of these two solutions are, respectively, given by:
\begin{align}\label{range_q_b}
q_{+}\equiv\cos\gamma&<q_{\rm bounce}(t)<\infty~,
\\
\label{range_q_ec}
q_{-}\equiv\sin\gamma&>q_{\rm EC}(t)>0~.
\end{align}
The above inequalities illustrate clearly that the solutions $q_{\rm bounce}(t)$ and $q_{\rm EC}(t)$ are spatially separated, in the sense that the ranges of these functions do not overlap. The asymptotic behaviour for the desired wave function, shown in \ref{asym_behaviour_gen} in the quantum domain, however, seems to require a transition from a bouncing solution to a collapsing solution and vice versa. On the other hand our earlier discussions show that classically, in real time $t$, there exists no such solutions that can effect such a transition between $q_{\rm bounce}(t)$ and $q_{\rm EC}(t)$. How, then, can one study the quantum transition from a bouncing to a collapsing solution, semi-classically?

\begin{figure}[h]
\centering
	\begin{subfigure}[b]{0.49\textwidth}
		\centering\includegraphics[width=\textwidth]{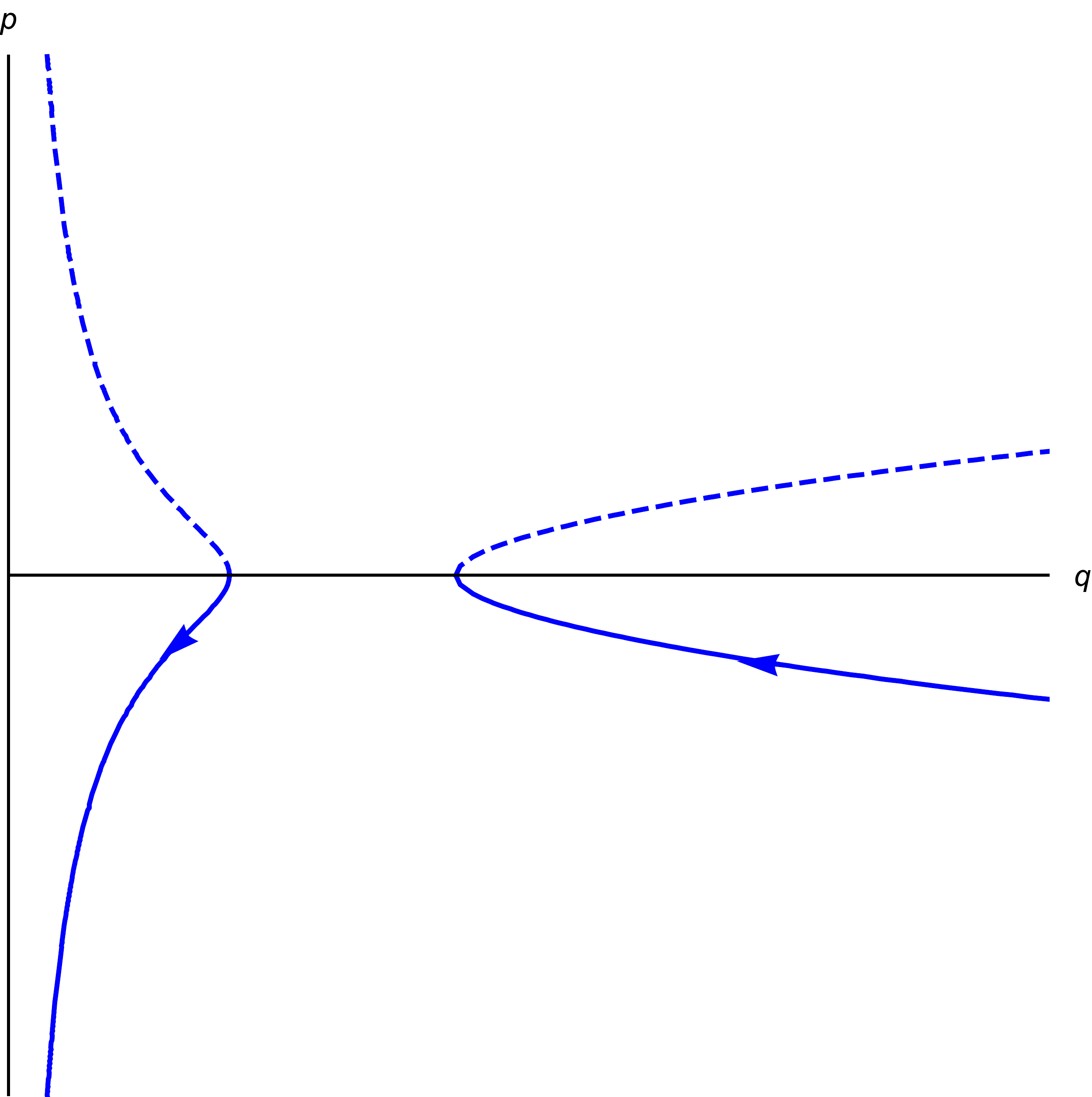}
		\caption{}
		\label{collapse_bounce_figure}
	\end{subfigure}
\hfill
\begin{subfigure}[b]{0.49\textwidth}
	\centering\includegraphics[width=\textwidth]{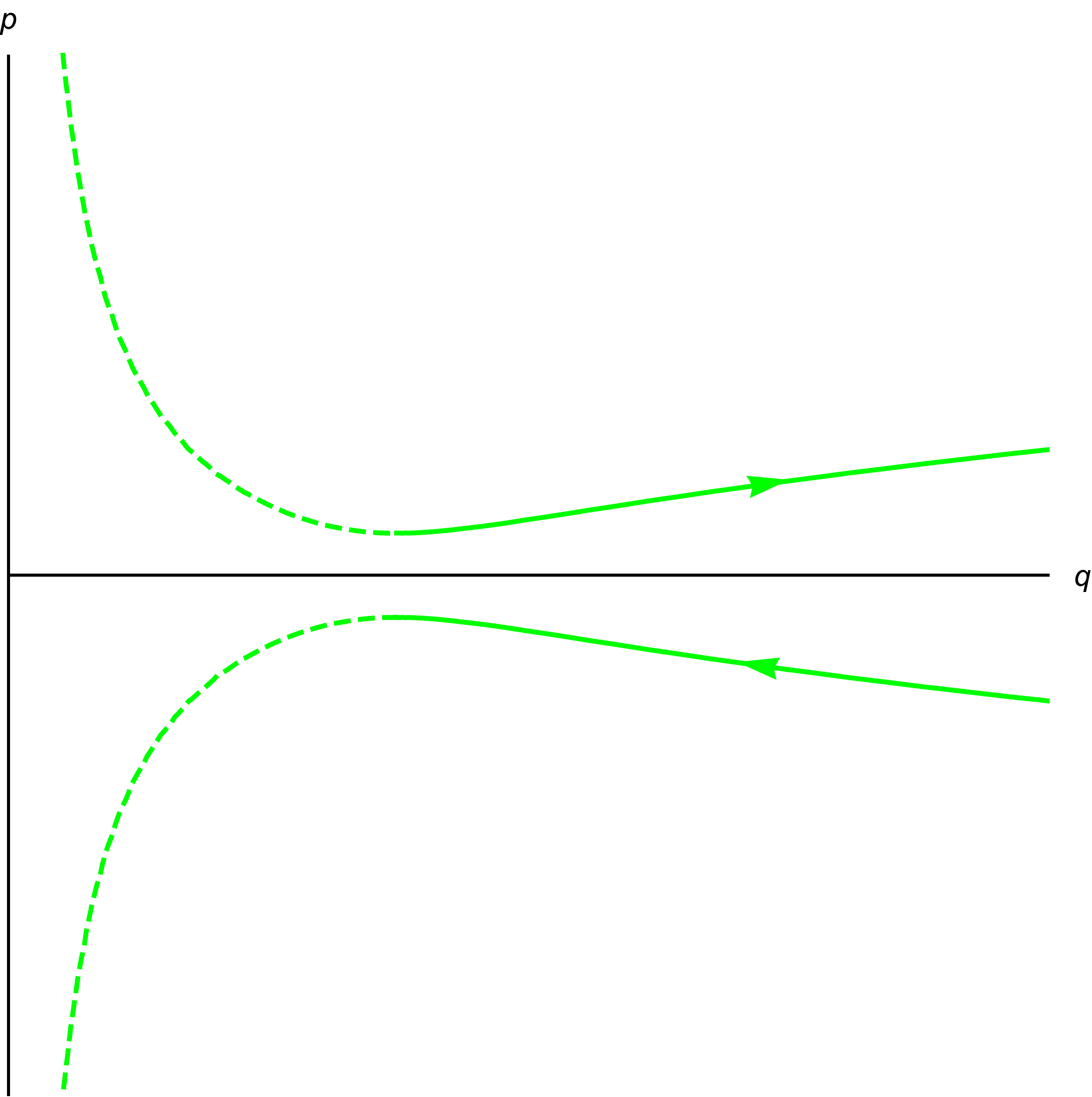}
	\caption{}
	\label{collapse_bounce_figure_2}
\end{subfigure}
\caption{(a) The semi-classical description in which the quantum mechanical tunnelling to singularity may be realized for $0<\rho_{\theta}/\rho_{0}<1/4$ must furnish a way of connecting the two disconnected continuous blue curves. (b) The semi-classical description in which a bounce can be realized, even when the shear density is such that $\rho_{\theta}/\rho_{0}>1/4$, must furnish a way of connecting the two disconnected green continuous curves. }
\end{figure}

\begin{figure}[h]
\centering
\begin{subfigure}[b]{0.48\textwidth}
	\includegraphics[width=\textwidth]{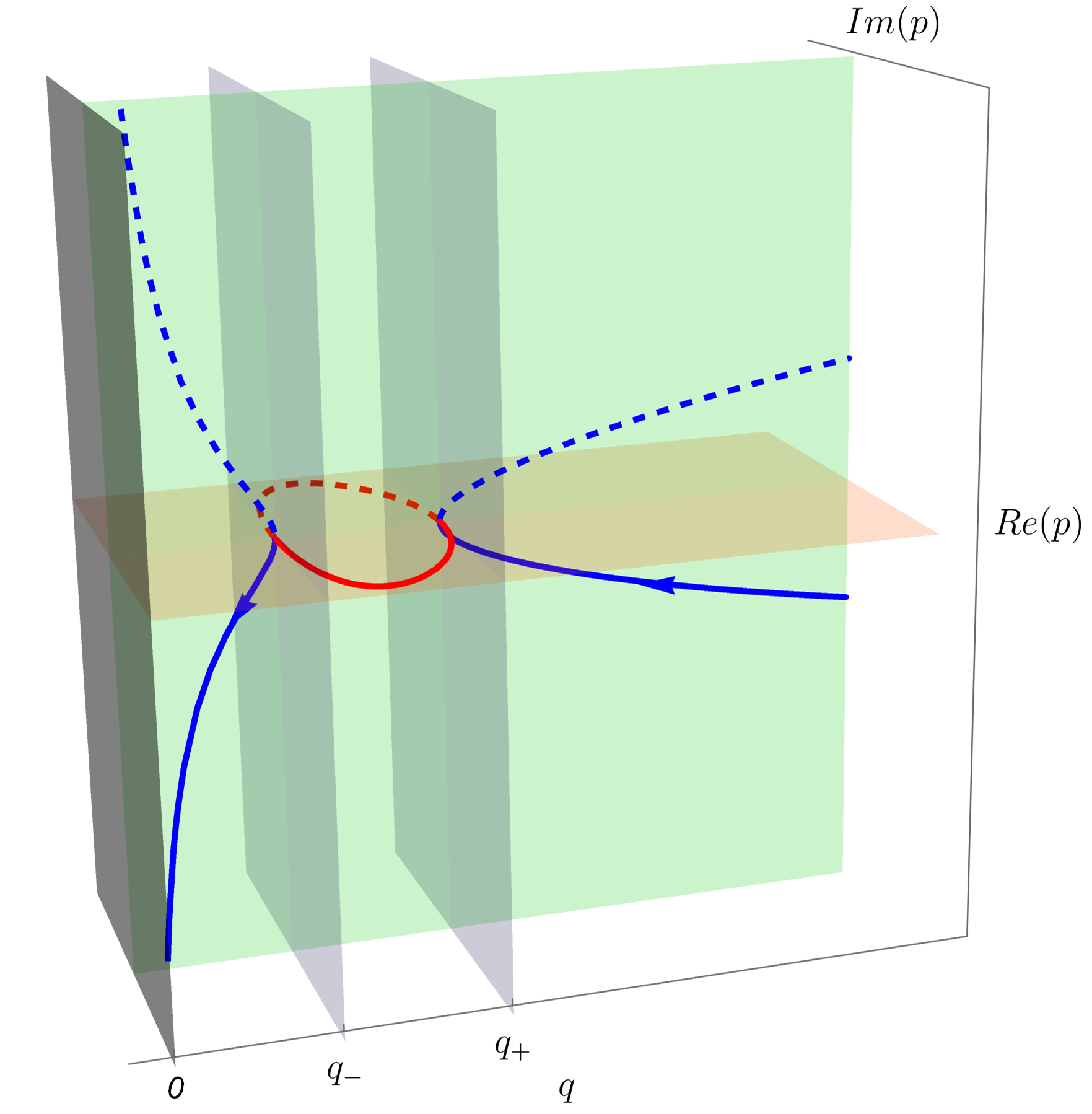}
	\caption{}
	\label{closed_euclid_curve_1}
\end{subfigure}
\hfill
\begin{subfigure}[b]{0.48\textwidth}
	\includegraphics[width=\textwidth]{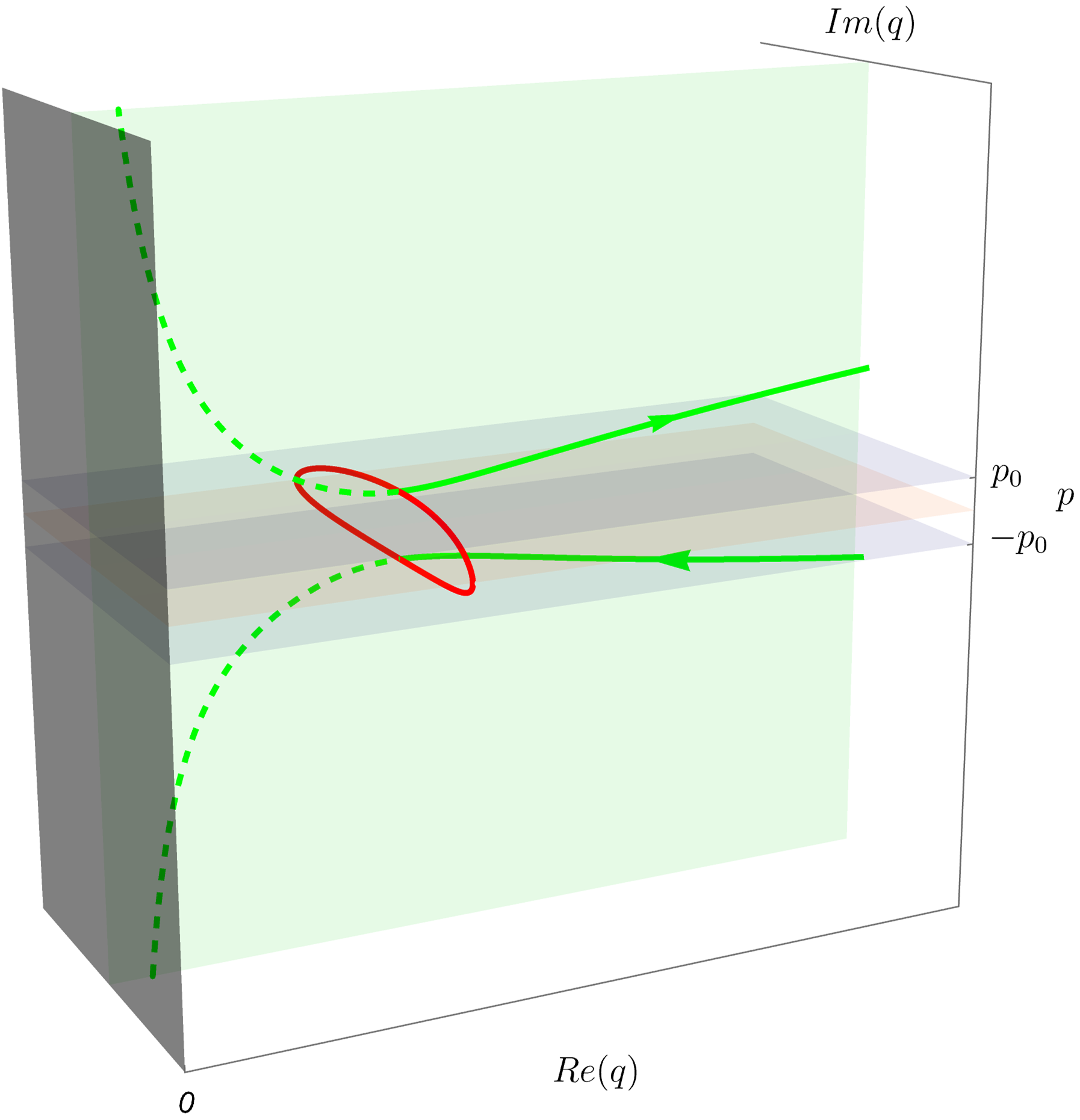}
	\caption{}
	\label{closed_euclid_curve_2}
\end{subfigure}
\caption{(a) $0<\rho_{\theta}/\rho_{0}<1/4$: The Euclidean solution (red curve) in \ref{Euclidean_sol} bridging the two disconnected classical solutions---one describing a bouncing scenario and the other a big crunch scenario---furnish a semi-classical interpretation for the process of tunnelling to the singularity. (b) $\rho_{\theta}/\rho_{0}>1/4$: The Euclidean solution (red curve) in \ref{Euclidean_sol_2} bridging the two disconnected classical solutions---one describing an ever expanding and the other an ever contracting universes---furnish a semi-classical interpretation for the process of bouncing, which is classically forbidden.}
\end{figure}

To answer this, we start by observing that under the complex time translation $t\rightarrow t\pm i(\pi/2h_{n}\mathcal{N})$, the solutions $q_{\rm bounce}(t)$ and $q_{\rm EC}(t)$ transform to each other. This motivates us to seek an Euclidean solution that bridges the two disconnected continuous curves in \ref{collapse_bounce_figure}. As we will see, in the semi-classical limit, going to the Euclidean sector furnishes the desired transition between a bouncing to a collapsing universe. To see this explicitly, note that under the Wick rotation, $t\rightarrow -it_{\rm E}$, the Hamiltonian constraint, presented in \ref{constraint_shear} transforms to:
\begin{align}\label{constraint_shear_eucl}
-\frac{M}{2\mathcal{N}^2}\left(\frac{dq}{dt_{\rm E}}\right)^{2}+\rho_{0}\left(1-q^{2}\right)-\frac{\rho_{\theta}}{q^2}=0
\end{align}
The momentum $p$ (which is defined as, $-(\partial L/\partial \dot{q})$, negative of the canonical momentum), on the other hand, transforms as $p\rightarrow ip_{\rm E}$, where $p_{\rm E}=\mathcal{N}^{-1}M\left(dq/dt_{\rm E}\right)$. Therefore, the Euclidean solutions may be imagined as contours of $H(p,q)=0$ in the $(q,{\rm Im}~p)$ plane. From \ref{solutions_b} and \ref{solutions_ec}, it follows that the explicit form of the Euclidean solution $q_{\rm Euc}(t_E)$ can be obtained by the Wick rotation, which yields,
\begin{align}\label{Euclidean_sol}
q_{\rm Euc}(t_{\rm E})&= \frac{1}{\sqrt{2}}\sqrt{1+\cos(2\gamma)\cos\left(2h_n\mathcal{N}t_{\rm E}\right)}~.
\end{align}
Clearly, $q_{\rm Euc}(t_{\rm E})$ is a periodic function of the Euclidean time $t_{\rm E}$, with the period being $T=\pi/(h_n\mathcal{N})$ and is described by a closed contour in the $(q,{\rm Im}~p)$ plane. Note that $q_-\leq q_{\rm Euc}(t_{\rm E})\leq q_+$, implying that the spacial size of the Euclidean universe has an upper and lower bound. This in turn, implies that the topology of, say, the $x-t_{E}$ cross-section of the Euclidean solution is a 2-torus. This Euclidean spacetime, along with the Lorentzian ones derived earlier, have been presented in \ref{closed_euclid_curve_1}, where the blue curves represent the Lorentzian solutions and the red closed curve represents the Euclidean solution. It is clear from \ref{closed_euclid_curve_1} that $q_{\rm Euc.}(t_{\rm E})$ acts as a bridge between the disconnected Lorentzian solutions and hence, describes tunnelling from a bouncing universe to a collapsing universe and vice versa. 

In order to derive the tunnelling amplitude we need to evaluate the on-shell value of the Euclidean action, whose exponentiation will yield the desired result. For that purpose, we start by applying the Wick rotation to the action $\mathcal{S}$ in \ref{action_ueff}, which now provides the Euclidean action $\mathcal{S}_{\rm E}$, with the following expression,
\begin{align}
\mathcal{S}_{\rm E}\equiv V_3\int dt_{\rm E}\left\{\frac{M}{2\mathcal{N}}\left(\frac{d q}{dt_{\rm E}}\right)^2-\frac{q^2}{2\mathcal{N}}\left[\left(\frac{d\theta_{1}}{dt_{\rm E}}\right)^2+\left(\frac{d \theta_2}{dt_{\rm E}}\right)^2\right]+\mathcal{N}\rho_{0}\left(1-q^{2}\right)\right\}~.
\end{align}
As in the Lorentzian case, in the Euclidean sector as well, the action $\mathcal{S}_{\rm E}$ is invariant under translation in the $(\theta_{1},\theta_{2})$ plane, which implies the following conservation relation,
\begin{align}\label{int_of_motion_E}
\mathcal{N}^{-1}q^2\frac{d\theta_i}{d t_{\rm E}}=-i\Pi_{i}\equiv \Pi_{{\rm E}~i}=\textrm{conserved}~;
\qquad i=1,2
\end{align}
Therefore the on-shell Euclidean action in terms of the Euclidean momenta $\Pi_{{\rm E}~i}$ takes the following form
\begin{align}
\mathcal{S}^{\rm on-shell}_{\rm E}\equiv V_{3}\int dt_{\rm E} \left[\frac{M}{2\mathcal{N}}\left(\frac{dq}{dt_{\rm E}}\right)^2-\frac{\mathcal{N}}{2}\frac{\Pi_{\rm E}^{2}}{q^{2}}+\mathcal{N}\rho_{0}\left(1-q^{2}\right)\right]~,
\end{align}
where, $\Pi_{\rm E}^{2}=\Pi_{{\rm E}~1}^{2}+\Pi_{{\rm E}~2}^{2}$. Variation of the above on-shell action with respect to the lapse function $\mathcal{N}$ and then substitution for the Euclidean momentum $p_{\rm E}$, yields the (Euclidean) constraint equation, which is given by,
\begin{align}\label{Energy_function_euclid}
\mathcal{H}_{\rm E}(p_{\rm E},q)\equiv-\frac{p_{\rm E}^2}{2M}+\rho_{0}\left(1-q^{2}\right)-\frac{\rho_{\theta}^{\rm E}}{q^2}=0~,
\end{align}
where, $\rho_{\theta}^{\rm E}=(1/2)\Pi_{\rm E}^{2}$. The above Hamiltonian constraint suggests that, the on-shell value of the action can be expressed as, $p_{\rm E}(dq/dt_{\rm E})$. Therefore, $\mathcal{S}^{\rm on-shell}_{\rm E}$ evaluated for the solution $q_{\rm E}(t_{\rm E})$, given by \ref{Euclidean_sol}, for one full rotation in the phase space, is given by:
\begin{align}
\mathcal{S}^{\rm on-shell}_{\rm E,(1)}&=\oint p_{\rm E}(q)dq
\nonumber
\\
&=\mathcal{A}_{(1)}=\frac{\pi\alpha_n\hbar}{2}(1-\sigma)~,
\end{align} 
where, $\mathcal{A}_{(1)}$ is the area enclosed by the Euclidean solution in the $(q,{\rm Im}~p)$ plane and the subscript $(1)$ indicates that the quantities are evaluated for one complete period of the Euclidean time $t_{\rm E}$. Following Landau and Lifschitz \cite{landau2013quantum}, we can evaluate the semi-classical ratio of the probability to bounce, denoted by $\mathcal{R}$ and that to collapse, which is denoted by $\mathcal{T}$, using the Euclidean action as:
\begin{align}\label{semi_class_TbyR}
\frac{\mathcal{T}}{\mathcal{R}}&\approx \exp\left[-\frac{\mathcal{S}^{\rm on-shell}_{\rm E,(1)}}{\hbar}\right]=\exp\left[\frac{\pi\alpha_n}{2}(\sigma-1)\right]~.
\end{align}
Using the above equation and the relation $\mathcal{R}+\mathcal{T}=1$, one can readily verify that the probability for collapse, $\mathcal{T}$, is given by \ref{T_approx}.

Let us now briefly consider the case in which a classical bouncing solution is not allowed. The shear density for this case falls in the range $(\rho_{\theta}/\rho_{0})>1/4$. However, as we have learned from the wave function $\Psi_{\vec{\Pi}}$ in \ref{WdW_section}, there is a quantum mechanical probability for the universe to bounce. In this parameter range, the two disconnected solutions are, respectively, the ever-expanding solution $q_{\rm E}(t)$ and the ever-contracting solution $q_{\rm C}(t)$. Hence, one expects that the semi-classical description of bounce may be realized by `bridging' these two disconnected solutions via a Euclidean instanton. This corresponds to connecting the continuous curves in \ref{collapse_bounce_figure_2}. Analogous to our discussion of the case $0<(\rho_{\theta}/\rho_{0})<1/4$, the Euclidean solution that enables this can be found by Wick rotation of the solutions $q_{E/C}(t)$, which gives: 
\begin{align}\label{Euclidean_sol_2}
q_{\rm Euc}(t)=\frac{1}{\sqrt{2}}\sqrt{1+i\sin\left(2 h_n\mathcal{N} t_E\right)\sinh(2\delta)}~.
\end{align}
Note that the above $q_{\rm Euc}(t)$ is also a periodic solution, but, unlike in the previous case (i.e., \ref{Euclidean_sol}), the corresponding Euclidean scale factor is explicitly complex. This can be seen visualized in \ref{collapse_bounce_figure_2}.  Proceeding as before, the semi-classical ratio of probability to bounce $\mathcal{R}$ and that to collapse $\mathcal{T}$ can be found from the Euclidean action as follows:
\begin{align}\label{semi_class_RbyT}
\frac{\mathcal{R}}{\mathcal{T}}&\approx \exp\left[-\frac{\mathcal{S}^{\rm on-shell}_{\rm E,(1)}}{\hbar}\right]=\exp\left[-\frac{\pi\alpha_n}{2}(\sigma-1)\right]~.
\end{align}
The expression for $\mathcal{R}$ and $\mathcal{T}$ that follows from the above, again matches exactly with \ref{T_approx}. The significance of the above result is that, enabled by the Euclidean spacetime, quantum mechanically one can realize a bouncing scenario even when the relevant parameters are such that this is classically forbidden. One can imagine this feature to extend beyond the simple model that we have focussed on this paper. In particular, if one starts with a universe consisting of only `normal matter' and the geometry evolving according to standard GR, we are eventually lead to an inevitable collapse to singularity. However, complex instanton geometries of the kind described in \ref{closed_euclid_curve_2} can enable quantum transition to an expanding universe. One can, presumably, go a step further and envision such a model which explains how our own Universe came into being.     

We shall now summarize this section. We have examined the semi-classical description of the non-trivial quantum mechanical features that arise out of the wave function $\Psi_{\vec{\Pi}}$, as derived in \ref{sol_psi_bar}, for our anisotropic Bianchi-I cosmological model. In the case where classical bounce is allowed and collapse to singularity is forbidden --- corresponding to $0<(\rho_{\theta}/\rho_{0})<1/4$ --- there exist a quantum mechanical probability $\mathcal{T}$ for tunnelling to singularity. This tunnelling, in turn, can be interpreted as being enabled by the Euclidean solution $q_{\rm Euc}(t_E)$ given in \ref{Euclidean_sol}. Moreover, the semi-classical value of the probability $\mathcal{T}$ can be calculated from the on-shell value of the Euclidean action and is found to match exactly with \ref{T_approx}. In case where classical bounce is forbidden --- corresponding to $(\rho_{\theta}/\rho_{0})>1/4$ ---there exist a quantum mechanical probability, denoted by $\mathcal{R}$ to bounce. This purely quantum process can be interpreted as being enabled by the Euclidean solution $q_{\rm Euc}(t_E)$ given in \ref{Euclidean_sol_2}, which is explicitly complex valued. In this case as well, the semi-classical value of the probability $\mathcal{R}$ that one calculates from the on-shell Euclidean action is consistent with \ref{T_approx}. Hence, in general, the quantum cosmological effects of our anisotropic Bianchi-I models can be ascribed to the existence of complex Euclidean spacetimes, leading to phenomenon, which are forbidden classically. In the next section, we shall find a path integral based approach, out of which these complex solutions emerge naturally. 
\section{Lorentzian path integral approach}\label{section_Lorentzian}

In recent years, several advances in the Lorentzian quantum cosmology (henceforth as LoQC, as a short hand notation) have introduced a new way of understanding the path integral approach to quantum cosmology. Before going into the application of the LoQC to the specific Bianchi I bouncing cosmology that we are focusing on currently, let us briefly review the essentials of this formalism. Towards this end, we shall closely follow the approach discussed in \cite{Feldbrugge:2017kzv,FeldbruggeJobLeon2019}. 

For convenience we have introduce the shorthand notation $\mathbf{Q}=(q,\theta_{1},\theta_{2})$, such that the solutions $\Psi(\mathbf{Q})$ to the Wheeler-deWitt equation can be generated via the following path integral:
\begin{align}\label{Psi_definition_0}
\Psi(\mathbf{Q})=\int_{-\infty}^{\infty}d\mathcal{N}\int_{0}^{\infty}dq'\int_{\mathbb{R}^2}d^2\vec{\theta}'\left[\int_{\mathbf{Q}'}^{\mathbf{Q}}\mathcal{D}[\mathbf{Q}'']\exp\left(\frac{i}{\hbar}\int_{0}^{1}\mathcal{L}[\mathcal{N},\mathbf{Q}'',\dot{\mathbf{Q}''}]dt\right)\right]\psi_{0}(\mathbf{Q}')~,
\end{align}
where, $\psi_{0}(\mathbf{Q})$ is an arbitrary `initial wave function' that we shall call the `seed' following \cite{FeldbruggeJobLeon2019}, $\mathcal{L}$ is the appropriate Lagrangian describing the classical dynamics in the mini-superspace, which in our case corresponds to the one presented in \ref{action_ueff} and $\mathcal{N}$ is the lapse function, gauged to satisfy $\dot{\mathcal{N}}=0$. By defining a `physical time' parameter $T=\mathcal{N}t$, we can recast the above path integral in the following form:
\begin{align}\label{Psi_definition}
\Psi(\mathbf{Q})=\int_{-\infty}^{\infty}dT\int_{0}^{\infty}dq'\int_{\mathbb{R}^2}d^2\vec{\theta}'\,\,\mathcal{G}_{\rm full}(\mathbf{Q},\mathbf{Q}';T-T_0)\psi_{0}(\mathbf{Q}')
\end{align} 
where, the function $\mathcal{G}_{\rm full}(\mathbf{Q},\mathbf{Q}';T-T_0)$, which we shall refer to as the `propagator', is defined as:
\begin{align}
\mathcal{G}_{\rm full}(\mathbf{Q},\mathbf{Q}';T-T_0)=\int_{\mathbf{Q}'}^{\mathbf{Q}}\mathcal{D}[\mathbf{Q}]\exp\left(\frac{i}{\hbar}\int_{T_0}^{T}\mathcal{L}[\mathcal{N}=1,\mathbf{Q},\dot{\mathbf{Q}}]dt\right)
\end{align}
The above path integral can be interpreted as defining the quantum mechanical propagator for a point particle moving in a fictitious three dimensional space\footnote{Strictly speaking, the fictitious point particle is moving in a three dimensional spacetime, as opposed to space, owing to $q$ being a time-like `direction'. We have, in fact, discussed this following \ref{metric_mss}. However, since this distinction does not significantly affect the forthcoming discussions, one can, for the most part, imagine $\mathcal{Q}$ to be the coordinates of a three-dimensional space.}, with generalized coordinates $\mathbf{Q}$ and whose dynamics is governed by the Lagrangian $\mathcal{L}[\mathcal{N}=1,\mathbf{Q},\dot{\mathbf{Q}}]$. Hence, in Dirac's bra-ket notation, \ref{Psi_definition} takes the following suggestive form:
\begin{align}\label{Psi_definition_2}
\ket{\Psi}=\int_{-\infty}^{\infty}e^{-\frac{i}{\hbar}\hat{\mathcal{H}}(T-T_0)}\ket{\psi_0}\,dT
\end{align}
where $\hat{\mathcal{H}}$ is the Hamiltonian operator corresponding to the Lagrangian $\mathcal{L}[\mathcal{N}=1,\mathbf{Q},\dot{\mathbf{Q}}]$, $\Psi(\bm{Q})\equiv \langle \bm{Q}\ket{\Psi}$ and $\mathcal{G}(\mathbf{Q},\mathbf{Q}';T-T_0)=\langle \bm{Q}|\exp[-(i/\hbar)\hat{\mathcal{H}}(T-T_0)]\ket{\bm{Q}'}$. Note that the Wheeler-deWitt equation in the present context reduces to $\hat{\mathcal{H}}\ket{\Psi}=0$. In summary, given the two ingredients --- (a) the propagator $\mathcal{G}$ (or, equivalently, the evolution operator depending on the Hamiltonian $\hat{\mathcal{H}}$) and (b) the seed $\psi_{0}$ --- a recipe for generating solutions to the Wheeler-deWitt equation is provided by \ref{Psi_definition} (or, equivalently, by  \ref{Psi_definition_2}). In what follows, we shall explicitly compute the relevant propagator and the seed that yields the wave function $\Psi_{\vec{\Pi}}(q,\vec{\theta})$ that we found in \ref{section_WdW_solution}. Following that, using Picard-Lefschetz theory, we shall demonstrate how the semi-classical description presented in \ref{WdW_semiclassical} emerge naturally from the exact path integral picture.
\subsection{Determination of the propagator} 

In this section, we will explicitly provide the expression for the propagator necessary to derive the solution of the Wheeler-deWitt equation. The first step in achieving the same corresponds to writing down the Lagrangian $\mathcal{L}[\mathcal{N}=1,\mathbf{Q},\dot{\mathbf{Q}}]$, which from \ref{action_ueff} takes the following form,
\begin{align}
\mathcal{L}[\mathcal{N}=1,\mathbf{Q}]=-\frac{1}{2}MV_{3}\dot{q}^2+\frac{V_{3}}{2}\left(\dot{\theta}^{2}_{1}+\dot{\theta}^{2}_{2}\right)q^{2}+\rho_{0}V_{3}\left(1-q^{2}\right)~,
\end{align}
where, the `mass' $M=(n-6)^{-2}(96/\kappa)$ and we have also used the expression for $U_{\rm eff}$ from \ref{intr_U_eff_II}. Path integral of the above Lagrangian will yield the propagator $\mathcal{G}_{\rm full}$. Since, there is translational symmetry along the $\vec{\theta}$ directions, it is more convenient to consider the reduced propagator $\mathcal{G}_{\vec{\Pi}}$, which is defined as follows:
\begin{align}\label{FT_propagator}
\mathcal{G}_{\rm full}(\mathbf{Q},\mathbf{Q}';T-T_0)=\int_{-\infty}^{\infty}\mathcal{G}_{\vec{\Pi}}(q,q';T-T_0)e^{i\frac{V_3}{\hbar}\vec{\Pi}.(\vec{\theta}-\vec{\theta}')}d^2\vec{\Pi}
\end{align} 
Direct substitution shows that $\mathcal{G}_{\vec{\Pi}}$ is the propagator corresponding to an one dimensional effective Lagrangian, which describes a particle interacting with an inverted harmonic oscillator potential ($\propto-q^2$) along with a singular $(1/q^{2})$ potential. The propagator for such a system can be computed using standard results available in the literature. However, for completeness, we have provided a brief derivation for the same in \ref{App_My}. From which the following explicit expression for the propagator can be read off:
\begin{align}\label{gen_kernel}
\mathcal{G}_{\vec{\Pi}}(q_2,q_1;T-T_0)&=\left(\frac{i\alpha_{n}\sqrt{q_2q_1}}{\sinh \left[h_{n}(T-T_{0}) \right]} \right)\exp \left[-\frac{i\alpha_{n}}{2}\coth \left[h_{n}(T-T_{0}) \right] \left(q_{1}^{2}+q_{2}^{2} \right)+\frac{i\alpha_{n}h_{n}\left(T-T_{0}\right)}{2} \right]
\nonumber
\\
&\hskip 2 cm \times I_{\nu}\left(\frac{i\alpha_{n}q_{1}q_{2}}{\sinh \left[h_{n}(T-T_{0}) \right]}\right)~,
\end{align}
where $I_{\nu}(z)$ denotes the modified Bessel function of the first kind and, the parameters $\alpha_{n}$ and $\nu$ are as defined in \ref{alpha_def_1} and \ref{def_mu2}, respectively. The `full' propagator $\mathcal{G}_{\rm full}$ can then be found using the definition \ref{FT_propagator}. However, as we shall shortly see, the relevant seed wave function $\psi_{0}(\mathbf{Q})$ for our problem turns out to be such that we only require the expression for $\mathcal{G}_{\vec{\Pi}}$.
\subsection{Emergence of the Wheeler-deWitt wave function}

We will present below the emergence of the Wheeler-deWitt wave function from an appropriate choice of the seed wave function $\psi_{0}(\bm{Q})$. Physically, the seed wave function $\psi_{0}(\bm{Q})$ encodes the `initial conditions' that are required to realize the path integral definition of a given solution to the Wheeler-deWitt equation. For instance, in the context of de Sitter cosmology, the original Hartle-Hawking no-boundary prescription is obtained by imposing the condition that the the universe should start from nothing, i.e., the universe should have a vanishing `initial size', in the path integral definition of wave function. Hence the origin of the name, no-boundary wave function \cite{Hartle:1983ai}. This condition, derived in the absence of anisotropy, in turn, can be achieved by requiring that $\psi_0(q)\propto\delta(q)$. However, recently, it has been demonstrated using LoQC that this definition leads to unsuppressed perturbations and hence to potential instability \cite{Feldbrugge:2017fcc,Feldbrugge:2017mbc,Feldbrugge:2018gin}. An alternative prescription, where one fixes the `initial' momentum conjugate to the scale factor to a specific Euclidean value, gets rid of this instability and realizes the Hartle-Hawking no-boundary wave function. Such an initial condition can be imposed by choosing the seed wave function to be of the form $\psi_{0}(q)\propto e^{-ik q}$, where $k$ is some specific imaginary number \cite{Lehners2019}. A similar choice can be made for defining the analogue of the no-boundary wave function in a class of bouncing models as well, as we have demonstrated in \cite{Rajeev:2021lqk}. 

However, unlike in the context of isotropic cosmological models, e.g., in de Sitter spacetime, the seed wave function in the present context of Bianchi-I cosmology depends on both the `scale factor' $q$ and the anisotropy parameters $\vec{\theta}$. Let us first explore the $\vec{\theta}$-dependence of the seed wave function $\psi_{0}(\mathbf{Q})$. Recall, from \ref{theta_momentum} that owing to the translational symmetry along $\vec{\theta}$ directions, $q^{2}(d\vec{\theta}/dT)=\vec{p}_{\theta}$ is a constant of motion. Thus, a sensible initial condition amounts to imposing a well defined initial value to the conjugate momentum  $\vec{p}_{\theta}$. Since the quantum mechanical representation of $\vec{p}_{\theta}$ is given by $-(i/\hbar)\partial_{\vec{\theta}}$, the suitable form of the seed wave function then reads:
\begin{align}
\psi_{0}(\mathbf{Q})=e^{i\frac{V_3}{\hbar}\vec{\Pi}.\vec{\theta}}\phi_{0}(q)~,
\end{align}
where, $\phi_{0}(q)$ is yet to be specified. We shall first present the explicit expression for $\phi_{0}(q)$ and then shall show that the solution to the Wheeler-deWitt equation indeed emerges. Subsequently, we will argue why the wave function $\phi_{0}(q)$ considered here is a reasonable choice for describing the wave function for an initially contracting Bianchi-I universe. We choose, except for some overall normalization constant:
\begin{align}\label{seed}
\phi_0(q)=\sqrt{q}\exp\left(-\frac{i M V_3h_n}{2\hbar}q^2\right)I_{\nu}\left(\frac{-ie^{-h_nT_0}MV_3h_nqq_0}{\hbar}\right)~,
\end{align}
where, $q_{0}$ and $T_{0}$ are constants. The foremost rationale behind the above choice of $\phi_0(q)$ is that it leads to the solutions $\Psi_{\vec{\Pi}}(\mathbf{Q})$ of the Wheeler-deWitt equation that we had obtained in \ref{section_WdW_solution}. In order to see this, let us use the above choice of the seed wave function in \ref{Psi_definition}. Using the expression for the propagator $\mathcal{G}_{\vec{\Pi}}$ from \ref{gen_kernel} and after performing an integration over both $\vec{\theta}'$ and $q'$, (see \ref{App_My} for details of the derivation) we arrive at the following expression for the wave function,
\begin{align}\label{Psi_T_integral}
\Psi_{\vec{\Pi}}(\mathbf{Q})=e^{i\frac{V_3}{\hbar}\vec{\Pi}.\vec{\theta}}\int_{-\infty}^{\infty}\mathcal{K}(q;T,T_{0})dT~,
\end{align}
where, the function $\mathcal{K}(q;T,T_{0})$ takes the form,
\begin{align}\label{seed_kernel}
\mathcal{K}(q;T)&=\sqrt{qq_{0}}e^{-h_{n}(T-T_{0})}\exp \left[\frac{i\alpha_{n}h_{n}\left(T-T_{0}\right)}{2} \right] 
\exp\left[-\frac{i\alpha_{n}}{2}q^{2}-\frac{i\alpha_{n}}{4}q_{0}^{2}\left(e^{-2h_{n}T}-e^{-2h_{n}T_{0}}\right)\right]
\nonumber
\\
&\hskip 1 cm \times I_{\nu}\left(-i\alpha_{n}qq_{0}e^{-h_{n}T} \right)~.
\end{align}
Now, making the substitution $s=(i\alpha_{n}q_0^2/4)e^{-2h_nT}$ and using the following integral representation for the Whittaker functions $M_{a,b}(z)$ in terms of the modified Bessel functions $I_{\nu}(z)$ (see, for instance, Eq.(13.16.4) of \cite{olver2010nist}):
\begin{align}
\frac{1}{\Gamma(1+2b)}M_{a,b}(z)=\frac{\sqrt{z}e^{-\frac{1}{2}z}}{\Gamma\left(\frac{1}{2}+b-a\right)}\int_{0}^{\infty}e^{-s}s^{-a-\frac{1}{2}}I_{2b}\left(2\sqrt{zs}\right)ds~;\qquad
\textrm{Re}[a-b]-\frac{1}{2}<0~,
\end{align}
it can be shown that the integral in \ref{Psi_T_integral} over the time coordinate $T$, evaluates to:
\begin{align}\label{Psi_Pi_expression}
\Psi_{\vec{\Pi}}(\mathbf{Q})\propto e^{i\frac{V_3}{\hbar}\vec{\Pi}.\vec{\theta}}z^{-\frac{1}{4}}M_{\mu,\nu}(z)~,
\end{align} 
where, $z=i\alpha_{n}q^{2}$ and, the parameters $\mu$ and $\nu$ are as defined in \ref{def_mu1} and \ref{def_mu2}, respectively. Note that the above expression matches exactly with that for $\Psi_{\vec{\Pi}}(\mathbf{Q})$, as we obtained in \ref{section_WdW_solution}, for instance, see \ref{sol_psi_bar}. Although the fact that \ref{Psi_T_integral} leads to \ref{Psi_Pi_expression} is a perfectly reasonable \textit{mathematical} justification for our choice of the seed wave function $\psi_{0}(\bm{Q})$, a satisfactory \textit{physical} justification is in order. We shall now address the same.
\subsection{Physical justification of the seed wave function}

We start by observing that the effect of the seed wave function $\psi_{0}(\bm{Q})$ in \ref{Psi_definition_0} can also be interpreted as the addition of a boundary term $\mathcal{S}_{\rm B}[\mathbf{Q}(T_0)]\equiv -i\hbar\log[\psi_{0}(\mathbf{Q}(T_0))] $ to the relevant action, so that the wave function becomes,
\begin{align}\label{Psi_definition_3}
\Psi(\mathbf{Q})=\int_{-\infty}^{\infty}dT\int_{0}^{\infty} dq'\int_{\mathbb{R}^{2}} d^{2}\theta'\int^{\mathbf{Q}(T)=\mathbf{Q}}_{\bm{Q}(T_{0})=\bm{Q}'}\mathcal{D}[\mathbf{Q}]~\exp\left[\frac{i}{\hbar}\left(\mathcal{S}+\mathcal{S}_{\rm B}\right)\right]~.
\end{align}  
The boundary term $\mathcal{S}_{\rm B}$, in turn, has the effect of imposing the following initial conditions onto the saddle point configurations that contribute to the above path integral,
\begin{align}
V_3\left.q^2\frac{d\vec{\theta}}{dT}\right|_{T_0}&=\lim_{\hbar\rightarrow 0}\frac{\partial \mathcal{S}_{B}[\mathbf{Q}(T_0)]}{\partial \vec{\theta}(T_0)}~,
\label{in_cond_theta}
\\
\left.-MV_3\frac{dq}{dT}\right|_{T_0}&=\lim_{\hbar\rightarrow 0}\frac{\partial \mathcal{S}_{B}[\mathbf{Q}(T_0)]}{\partial q(T_0)}~.
\label{in_cond_q}
\end{align}
Since the right hand side of the above conditions explicitly involve the limit $\hbar \rightarrow 0$, it follows that the expression for $\mathcal{S}_{\rm B}$ to leading order in $\hbar$ determines the initial conditions imposed on the saddle point configurations. The leading order behaviour of $\mathcal{S}_{\rm B}$ in the $\hbar \rightarrow 0$ limit can be found from the known uniform asymptotic expansion for the modified Bessel functions (see, Eq. (10.41.3) of \cite{olver2010nist}) and it turns out to be:
\begin{align}\label{S_B_semiclassical}
\frac{\mathcal{S}_{\rm B}}{V_3}&\approx\vec{\Pi}\cdot\vec{\theta}(T_0)-\frac{h_n M }{4}\left[q_0^2 e^{-2 h_n T_0 }+2 q^2(T_0)\right]
\nonumber
\\
&\hskip 2 cm +\left(\frac{M h_n\sigma}{2}\right)\left[\sqrt{1+\zeta^2(q(T_0),T_0)}+\log\left(\frac{\zeta(q(T_{0}),T_0)}{1+\sqrt{1+\zeta^2(q(T_0),T_0)}}\right)\right]~.
\end{align}
where,
\begin{align}
\zeta(q,T)&=\frac{2}{\sigma}e^{-h_nT}qq_0~,
\end{align}
and $q_{0}$ is the arbitrary constant appearing in the seed wave function. From \ref{in_cond_theta} and the expansion of the boundary action, presented in \ref{S_B_semiclassical}, it is immediate that the boundary action $\mathcal{S}_{\rm B}$ imposes the following initial conditions on the $\vec{\theta}$, i.e., anisotropic degrees of freedom:
\begin{align}
q^{2}(T_0)\dot{\vec{\theta}}(T_0)=\vec{\Pi}~.
\end{align} 
These are clearly the desired ones, as the momenta conjugate to the anisotropic degrees of freedom got fixed by this boundary condition. The initial condition on the $q$ degree of freedom is more subtle. To extract this conveniently, we first make the reasonable assumption that the initial time $T_0$ is far in the asymptotic past, i.e., $-(h_nT_0)\gg 1$. In this limit, the boundary condition, as presented in \ref{in_cond_q} reduces to the simplified form,
\begin{align}
\dot{q}(T_0)-h_nq(T_0)\approx -h_nq_0e^{-h_nT_0}<0~.
\end{align}
The above condition was shown to unambiguously dispense the classical solution of a particle in one-dimensional inverted harmonic oscillator potential, corresponding to a particle directed towards the left \cite{Rajeev:2021zae}. In the context of mini-superspace model of Bianchi-I cosmology, when $T_0$ is far enough in the asymptotic past, an initially contracting classical solution will have a sufficiently large value of $q$, such that the effect of the $(1/q^{2})$ term in the potential in \ref{WdW_II} may be negligible. Therefore, in this limit, the effective dynamics of the $q$ degree of freedom is approximately that of a particle in an inverted harmonic oscillator potential. Therefore, the aforementioned initial condition of `incoming particle from the right' for inverted harmonic oscillator potential becomes directly applicable in the present context, and translates to a universe that is initially contracting. Another way of seeing this is to consider first the semi classical limit of $\phi_0(q)$, followed by the limit $-(h_nT_0)\gg 1$, which yields:
\begin{align}
\phi_0(q)\approx \exp\left[\frac{iMh_n}{2\hbar}qq_0e^{-h_nT_0}+\textrm{terms~independent~of~}T_{0}+\mathcal{O}\left(e^{h_{n}T_{0}}\right)\right]~.
\end{align}
Clearly, for sufficiently large negative value of $T_0$, the above wave function corresponds to $\phi_{0}\sim \exp(i\gamma q)$, where $\gamma$ is independent of $q$. Then, $p\phi_{0}=-i(\partial \phi_{0}/\partial q)=\gamma \phi_{0}$, thus from the result $p\propto -\dot{q}$, it follows that for the universe described by $\phi_{0}(q)$, the `scale factor decreases with time, as $\dot{q}$ is a negative quantity and hence it does describe a contracting universe. Stated alternatively, we have just substantiated our claim that the solution $\Psi_{\vec{\Pi}}$ corresponds to a universe that is initially contracting. This completes the physical justification for our choice of the seed function $\psi_{0}(q)$.  
\subsection{The Picard-Lefschetz theory and the emergence of instantons}
 
We shall now examine the semi-classical limit of the path integral approach that we have elaborated in the previous subsections. As evident from \ref{alpha_def_1}, the semi-classical limit, which corresponds to $\hbar\rightarrow 0$, can also be considered as the $\alpha_{n}\gg 1$ limit. In particular, we will illustrate how the instanton spacetimes, described  in a rather ad-hoc manner in \ref{WdW_semiclassical}, naturally emerge from the semi-classical limit of the path integral definition of $\Psi_{\vec{\Pi}}$, say, as give in \ref{Psi_definition_0}. 

For this purpose, we start with \ref{Psi_T_integral}, which follows from \ref{Psi_definition_0} after substituting appropriate expression for the propagator $\mathcal{G}$ and a suitable choice of the seed wave function $\psi_{0}$, as described in \ref{seed}. The semi-classical limit of $\Psi_{\vec{\Pi}}$ can then be investigated by taking $\alpha_n\gg 1$ on the right-hand side of \ref{Psi_T_integral}. By making use of the standard uniform asymptotic expansion of modified Bessel functions, one arrives at:
\begin{align}\label{Psi_S_int}
\Psi_{\vec{\Pi}}(q,\vec{\theta})\approx e^{i\frac{V_3}{\hbar}\vec{\Pi}.\vec{\theta}}\int_{-\infty}^{\infty}\mathcal{F}(q,T)\exp\left[i\alpha_n\tilde{S}(q,T)\right]dT~;\qquad
\alpha_{n}\gg 1~,
\end{align}
where, the `pre-factor' $\mathcal{F}$ and the `effective action' $\tilde{S}$ are as defined below:
\begin{align}
\mathcal{F}(q,t)&\propto \frac{\sqrt{q}}{\left[1+\zeta^2(q,T)\right]^{1/4}}e^{-h_n T}~,
\\
\tilde{S}(q,T)&=-\frac{1}{4}\left(-2 h_n T +q_0^2 e^{-2 h_n T }+2 q^2\right)
\\
\nonumber
&+\left(\frac{\sigma}{2}\right)\left[\sqrt{1+\zeta^2(q(T),T)}+\log\left(\frac{\zeta(q,T)}{1+\sqrt{1+\zeta^2(q(T),T)}}\right)\right]~.
\end{align}
Note that, we have taken out a factor of $\alpha_{n}$ to define $\tilde{\mathcal{S}}$, to better illuminate that the semi-classical limit: $\alpha_{n}\gg 1$, can be effected via saddle point approximation of \ref{Psi_S_int}. The physical interpretation of the action $\tilde{S}(q_1,T)$ is the following: $\alpha_{n}\tilde{S}(q_1,T)+V_3\vec{\Pi}.\vec{\theta}_1$ gives us the action $\mathcal{S}+\mathcal{S}_{B}$, evaluated at a classical configuration of $\mathbf{Q}$, that satisfy the boundary conditions $\mathbf{Q}(t=T)=\mathbf{Q}_1$, along with \ref{in_cond_theta} and \ref{in_cond_q}. Let us now look at how the instanton spacetimes emerge from the saddle point approximation of \ref{Psi_S_int}.   

Despite the rather complicated expression for $\tilde{S}$, the saddle points, defined by $\partial_{T}\tilde{S}=0$, can be computed explicitly to get:
\begin{align}\label{Saddle_soln}
h_{n}T_{j,\pm}=\frac{1}{2} \log \left[\left(\frac{q_0^2 }{\sigma ^2-1}\right)\left(1-2 q^2\pm\sqrt{\left(2 q^2-1\right)^2+\sigma ^2-1}\right)\right]+ i\pi j~;\quad j\in\mathbb{Z}~.
\end{align}
Although a formal procedure to solve for the saddle points yield the above expression, some of $h_n T_{j,\pm}$ may not actually be saddle points at all. However, for convenience, we shall take \ref{Saddle_soln} as just the definition of the quantities $h_n T_{j,\pm}$, which may or may not correspond to a saddle point of $\tilde{S}$, depending on the value of the parameters. Nevertheless, for any value of parameters, there are an infinite number of saddle points. But, not all of them contribute to the saddle point approximation of \ref{Psi_S_int}. Picard-Lefschetz theory can be applied to determine the relevant saddle points for such oscillatory integrals. In this section, however, our main goal is to show that some of complex saddle points can become relevant and they correspond to the Eulidean instanton spacetimes that we considered in \ref{WdW_semiclassical}. Moreover, we shall also show that the imaginary part of effective action, when evaluated for relevant complex saddle points, leads to the semi-classical amplitudes $\mathcal{R}$ and $\mathcal{T}$ that we had obtained therein. 
\begin{figure}[h]
	\centering
	\begin{subfigure}[b]{0.49\textwidth}
		\includegraphics[width=\textwidth]{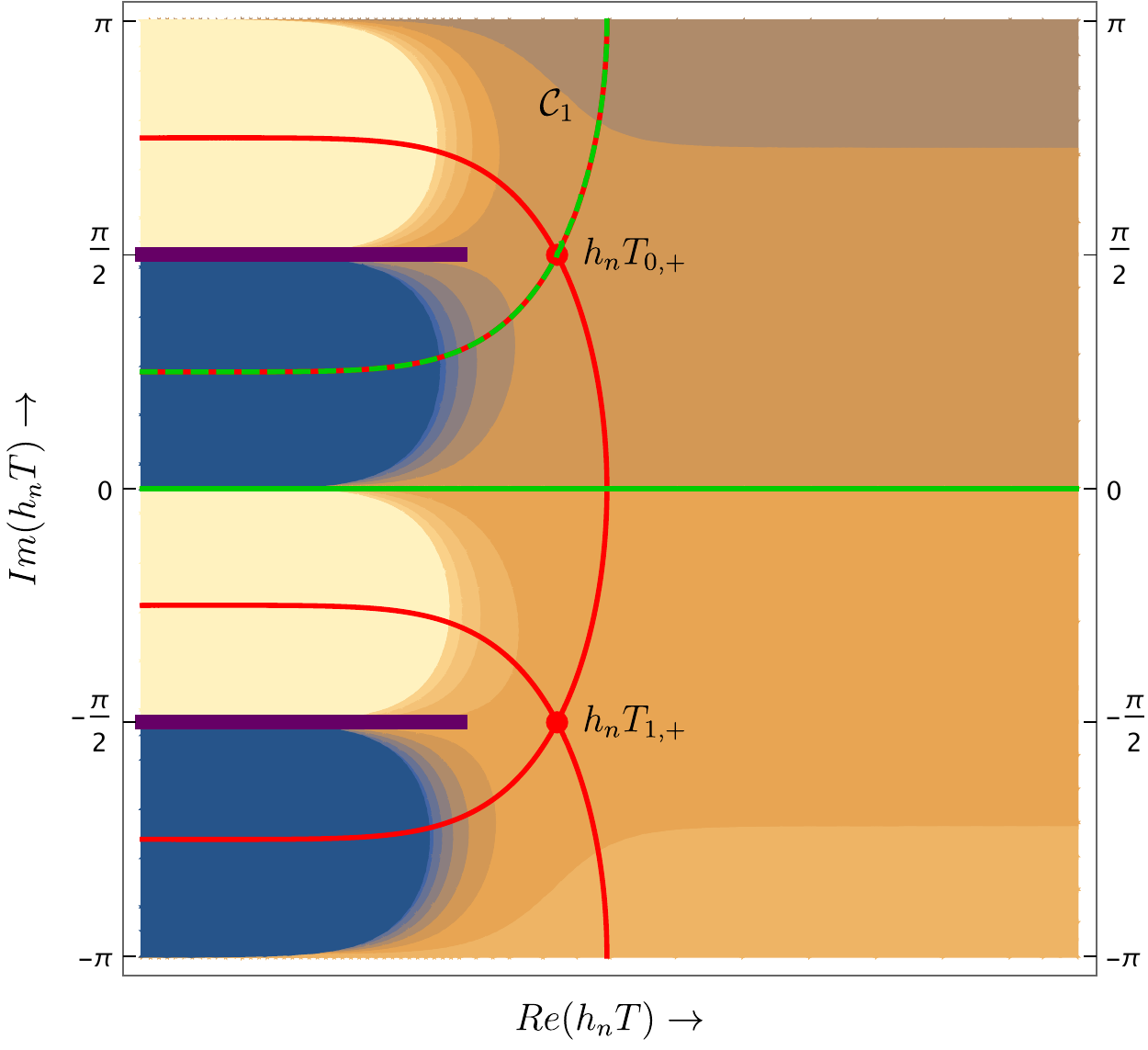}
		\caption{}
		\label{Picard_Lefschetz_tunnel}
	\end{subfigure}
	\hfill
	\begin{subfigure}[b]{0.49\textwidth}
		\includegraphics[width=\textwidth]{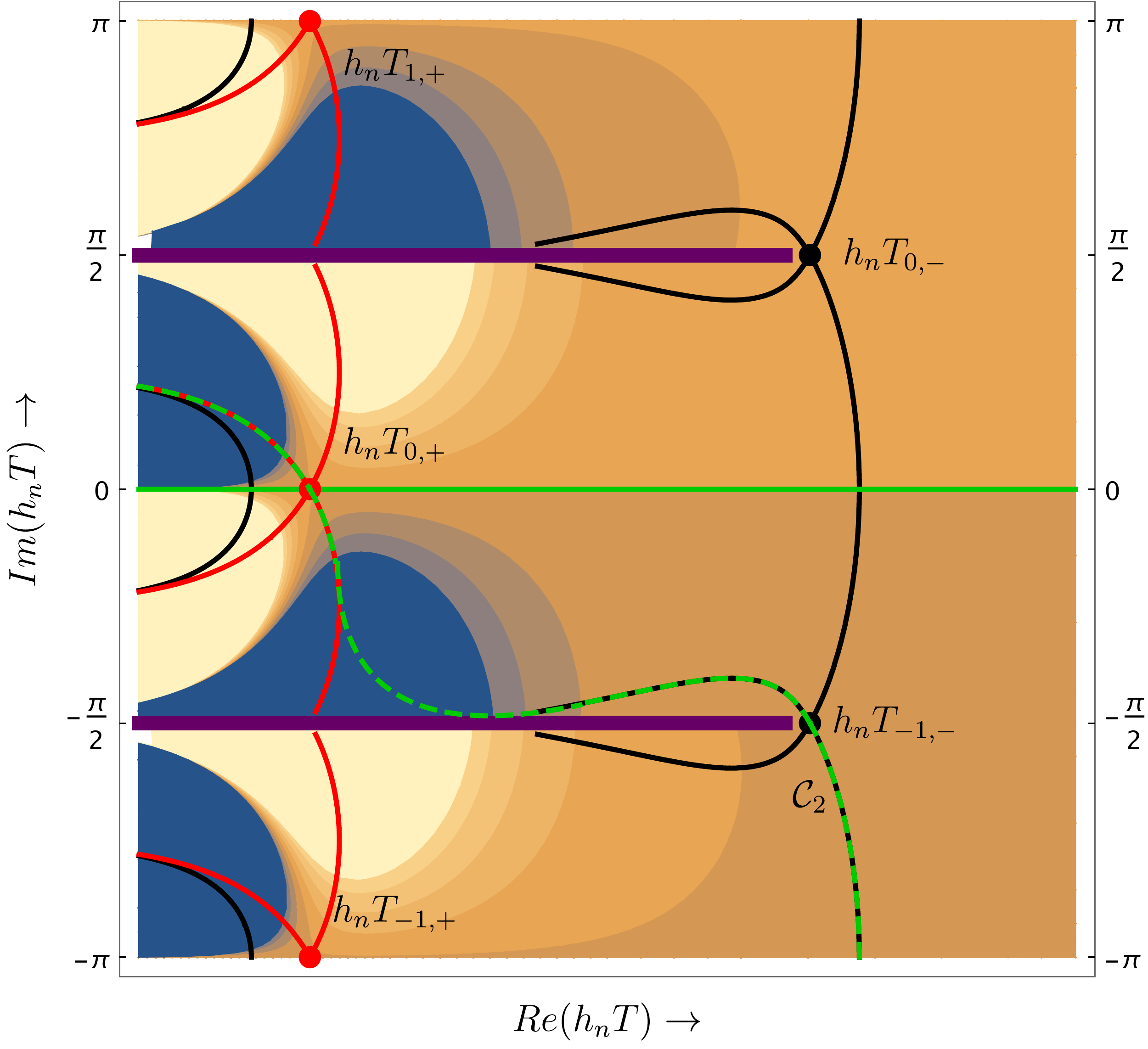}
		\caption{}
		\label{Picard_Lefschetz_bounce}
	\end{subfigure}
\caption{(a) \textbf{Case 1:} $0<\sigma<1$. The non-trivial quantum feature, in this case, arising as a result of the complex saddle point $T_{0,+}$ and corresponds to the Euclidean spacetime shown in \ref{collapse_bounce_figure}. (Plot is generated for $q=0.1,q_0=1,\sigma=.5,h_n=1$.) (b) \textbf{Case 2:} $\sigma>1$. There are two relevant saddle points in this case, namely,$T_{0,+}$ and $T_{-1,-}$. The former corresponds to the classical scenario of an ever contracting universe, while the latter is responsible for possibility of quantum bouncing. The complex saddle point $T_{-1,-}$ corresponds to the Euclidean spacetime shown in \ref{collapse_bounce_figure_2}. (Plot is generated for $q=10,q_0=1,\sigma=1.5,h_n=1$.) }
\label{Picard_Lefschetz}
\end{figure}

To facilitate the application of the Picard-Lefschetz theory, in \ref{Picard_Lefschetz}, we have presented the required steepest ascent and descent contours of $\mathcal{S}$, that emanate from the saddle points. A convenient graphical convention has been used in these figure, which is summarized below:
\begin{itemize}

\item The horizontal continuous green curve, described by the equation $\textrm{Im}(T)=0$, is the original integration contour in \ref{Psi_S_int}.

\item Thick horizontal purple half-lines represent the branch-cuts of $\tilde{S}$.

\item Red and black circular blobs represent the saddle points $T_{j,+}$ and $T_{j,-}$, respectively. The continuous curves emanating from these blobs represent the steepest descent/ascent curves.

\item To better visualize the ascent/descent directions, we have also presented the coloured contours of $\textrm{Re}[\tilde{S}]$. The colour coding is as follows: darker shades of blue represent more negative values, while lighter shades of orange correspond to more positive values. 

\item The contour to which the original horizontal contour should be deformed, to enable saddle point approximation, is represented by dashed green curve.

\end{itemize} 
We will now use these steepest descent/ascent contours along with the associated saddle points to work out the semi-classical limit explicitly. 

\subsubsection{Semi-classical limit with sub-dominant shear density}

In this section we will discuss the scenario in which the shear density $\rho_{\theta}$ is sub-dominant, i.e., it satisfies the condition, $0<\rho_{\theta}<(\rho_{0}/4)$ (or, equivalently, $0<\sigma<1$). This corresponds to the case, where bounce is classically allowed. Therefore, a universe that starts off contracting from a large enough initial size cannot reach the singularity, classically. But, as we have already seen in \ref{WdW_section}, there is a quantum mechanical amplitude for tunnelling to zero size. Since the non-trivial quantum features in this case arise near the singularity, located at $q=0$, we must consider the saddle points of the `effective action' $\tilde{S}$ for $q\ll1$. It turns out that, in this range of parameters, the only relevant saddle points are given by $h_{n}T_{+,j}$, where $j\in\mathbb{Z}$. When $0<\sigma<1$, \ref{Saddle_soln} tells us that $h_{n}T_{+,j}$ is complex for all values of $j\in\mathbb{Z}$. This is a reflection of the fact that regions near singularity is forbidden classically and the solution cannot acquire values $q(t)\ll 1$ for any real value of time $t$. 

Now, Picard-Lefschetz theory dictates that relevant saddle points are only those whose steepest ascent contours intersect the original integration contour, which is $\textrm{Im}(T)=0$. From \ref{Picard_Lefschetz_tunnel}, we find that the only relevant saddle point is given by $h_nT_{0,+}$, which has a positive imaginary component, $\textrm{Im}(h_nT_{0,+})=\pi/2$. One can now obtain an absolutely convergent integral representation of $\Psi_{\vec{\Pi}}$ by deforming the integration contour to $\mathcal{C}_1$, as shown in \ref{Picard_Lefschetz_tunnel}, which passes through the relevant saddle point $T_{0,+}$. The physical implication of $T_{0,+}$ being complex is that a classical universe, satisfying the constraint equation, can contract to near-zero size after evolving along imaginary time for a duration $\delta t$, which is given by $\delta t=\textrm{Im}(T_{0,+})=(i\pi/2h_n)$. This is precisely the manner in which the Euclidean solution in \ref{Euclidean_sol} was defined as a bridge between the two disconnected continuous curves in \ref{collapse_bounce_figure}. Let us now also look at the action $\tilde{S}$ evaluated at the relevant saddle point $T_{0,+}$. In particular, the imaginary component of the on-shell action gives:
\begin{align}
\textrm{Im}[\alpha_n\tilde{S}(q,T_{0,+})]=\frac{\pi\alpha_n}{4}(\sigma-1)~.
\end{align}
Therefore, the semi-classical probability for tunnelling to singularity, relative to that for bounce, is given by
\begin{align}
\frac{\mathcal{T}}{\mathcal{R}}&\approx\left|\exp\left(i\alpha_n\tilde{S}(q,T_{0,+})\right)\right|^{2}=\exp\left[\frac{\pi\alpha_n}{2}(\sigma-1)\right]~,
\end{align} 
which matches exactly with \ref{semi_class_TbyR}. We will now take up the case in which shear density dominates the energy budget of the universe.

\subsubsection{Semi-classical limit with dominant shear energy density}

In the case of dominant shear energy density, i.e., with $\rho_{\theta}>(\rho_{0}/4)$ the bounce is classically forbidden. This means that a universe that starts off contracting initially, inevitably reaches the singularity, classically. However, quantum mechanically there is a probability for the universe to bounce. Since, in this case, the non-trivial quantum phenomenon (i.e., bounce) is more evident for larger sizes of the universe, we shall fix $q\gg1$. For this range of parameters, we find that both $T_{j,\pm}$ are genuine saddle points of $\mathcal{S}$. In particular, $T_{0,+}$ is real, while $T_{j,-}$ is complex. Physically, therefore, $T_{0,+}$ corresponds to a classical universe that is monotonically contracting, while, $T_{j,-}$, potentially corresponds to the quantum bouncing scenario. 

To determine the relevant saddle points, we again resort to the Picard-Lefschetz theory. From \ref{Picard_Lefschetz_bounce}, we find that the relevant saddle points are located at $h_nT_{0,+}$ and $h_nT_{-1,-}$. An absolutely convergent integral representation for $\Psi_{\vec{\Pi}}$ can now be obtained by deforming the original integration contour in \ref{Psi_S_int} to the contour $\mathcal{C}_2$ in \ref{Picard_Lefschetz_bounce}. It is expected that $T_{0,+}$ is a relevant saddle point, as it describes an ever-contracting classical universe. The physical implications of the complex saddle point located at $h_nT_{-1,-}$ is that a classical universe, satisfying the constraint equation, which is ever contracting, can bounce after evolving along imaginary time for a duration of $\delta t=\textrm{Im}(T_{-1,-})=-(i\pi/2h_n)$. This is reminiscent of the Euclidean solution in \ref{Euclidean_sol_2}, which was interpreted as a bridge between the disconnected continuous curves in \ref{collapse_bounce_figure_2}. Let us now consider the action $\mathcal{S}$ evaluated at $T_{-1,-}$. In particular, the imaginary component of the on-shell action yields:
\begin{align}
\textrm{Im}\left[\alpha_{n}\tilde{\mathcal{S}}(q,T_{-1,-})\right]=\frac{\pi\alpha_{n}}{4}(1-\sigma)~.
\end{align}
Hence, the semi-classical probability for bouncing, relative to contracting towards the singularity, can be computed as:
\begin{align}
\frac{\mathcal{R}}{\mathcal{T}}&\approx\left|\exp\left(i\alpha_{n}\tilde{S}(q,T_{-1,-})\right)\right|^{2}=\exp\left[\frac{\pi\alpha_{n}}{2}(1-\sigma)\right]~.
\end{align}  
which matches precisely with \ref{semi_class_RbyT}. To summarize, we have illustrated that Picard-Lefschetz theory not only offers us a method in which the instanton spacetimes, that enable quantum tunnelling and quantum bouncing, emerge naturally but also provides us a more rigorous way to derive the semi-classical value of $\mathcal{T}$ and $\mathcal{R}$ from a path integral perspective.
        
\section{Discussion}\label{section_discussion}

Bouncing models replace the general relativistic singularity of standard Big-Bang cosmology with a cosmic bounce, which is a regular phase of transition from a contracting to an expanding universe. As a result, the horizon problem and flatness problem that plague the standard Big-Bang cosmology gets resolved without the introduction of inflation. However, on the downside, the contracting phase leads to an undesirable growth of anisotropy in the universe. Different proposals have been put forward to address this pathology in bouncing models. In view of this, we have addressed the following question in the present work: Can the instabilities of a bouncing universe due to anisotropies get magnified or reduced as a result of quantum cosmological effects?  Since there are several different types of bouncing models, with the motivations and physical content varying over a wide range, it would be difficult, if not impossible, to address this question in its full generality. Therefore, to make progress, we have considered a family of exactly solvable anisotropic bouncing models, whose stress $\rho_{\theta}$ can be tuned to either allow a bounce or lead to a BLK-like instability. The finer details of how this specific model may be realized from a more microscopic theory is deliberately not addressed. Rather, we adopt a phenomenological approach in which the essential aspects of this model is assumed to be captured by an effective potential in the mini-superspace. 

The quantum cosmology of our model is then analyzed using two different approaches --- (i) using the Wheeler-deWitt equation and (ii) using the Lorentzian path integral formalism. As we have explicitly demonstrated, both of these approaches lead to a wave function $\Psi_{\vec{\Pi}}$ that reveal non-trivial quantum features, while relevance of the same to the present day cosmology is something we leave for the future. Firstly, when the shear density $\rho_{\theta}$ is tuned such that bounce is allowed classically, there is a non-zero (quantum) probability for the universe to collapse to singularity. On the other hand, when $\rho_{\theta}$ is such that bounce is classically forbidden and the universe is inevitably met with a BKL-like instability, the quantum cosmological effects facilitate the possibility of bouncing. As a result, one concludes that even when one has fine-tuned the parameters of the bouncing model to prevent a BKL-like instability, classically, there is a possibility for the universe to tunnel into zero size. Similarly, when the parameters are such that bounce is not classically allowed, which also turns out to be the case for a universe contracting with normal matter content, there is a possibility of quantum-enabled bounce. It is reasonable to imagine that this latter scenario is a feature of models that are more general than the simple one considered in this work. Therefore, perhaps, one can envision the possibility of our own Universe being an outcome of such a quantum bounce.        

The quantum origin of bounce, that we have just described, acquires interesting interpretations in the semi-classical analysis of the problem. This effect can be ascribed to the existence of complex Euclidean spacetimes that interpolate between different disconnected classical spacetimes. Moreover, the semi-classical amplitudes for the quantum processes can be found from the corresponding Euclidean action in the standard manner. The role of complex spacetimes in enabling quantum bounce is more transparent in the Lorentzian path integral approach. Therein, the application of the Picard-Lefschetz theory shows that, in the parameter ranges where quantum effects are significant, there exist relevant saddle points that are complex valued. These complex saddle points, in turn, can be naturally interpreted as corresponding to the complex Eulidean spacetimes that we mentioned before. 

It is reasonable to expect that the quantum effects that we have addressed in this work have important consequence to physical cosmology. In an upcoming paper we shall consider a more detailed analysis of this model, with the inclusion of realistic matter fields and perturbations. In particular, it is worth exploring, whether the quantum instabilities that arise due to tunnelling has any relevance to the stability of the perturbations. Similarly, the possibility of the quantum bounce, that ensues in a classically collapsing universe, as a viable model for our own Universe is also a promising avenue for further exploration. These we leave for the future.
\section*{Acknowledgements}

K.R. is supported by the Research Associateship of Indian Association for the Cultivation of Science (IACS), Kolkata, India. Research of S.C. is funded by the INSPIRE Faculty fellowship from the DST, Government of India (Reg. No. DST/INSPIRE/04/2018/000893) and by the Start-Up Research Grant from SERB, DST, Government of India (Reg. No. SRG/2020/000409). Research of V.M. is funded by the INSPIRE fellowship from the DST, Government of India (Reg. No. DST/INSPIRE/03/2019/001887). 
\section*{Dedication}

This article is dedicated to the memory of Prof. T. Padmanabhan, known to all of us as our beloved Paddy, who passed away on 17th September. Incidentally, Paddy started his monumental career with quantum cosmology and in particular, he had one of the initial papers on anisotropic quantum cosmology \cite{padmanabhan1981quantum}. May one of the brightest and fastest thinkers of our time rest in peace.
\appendix	
\labelformat{section}{Appendix #1}
\labelformat{subsection}{Appendix #1}
\labelformat{subsubsection}{Appendix #1}
\section{An Alternative solution to the Wheeler-deWitt equation}\label{App_Alt}

For completeness, we will present an alternative route to arrive at the solution of the Wheeler-deWitt equation derived in \ref{section_WdW_solution}. The starting point of this alternative approach is the Schr\"{o}dinger equation for an isotropic harmonic oscillator with an inverse quadratic potential in $N$ dimensions, which take the following form \cite{oyewumi2003isotropic},
\begin{align}
R''(r)+\left(\frac{N-1}{r}\right)R'(r)-\frac{\ell(\ell+N-2)}{r^{2}}R(r)+\frac{2m}{\hbar^{2}}\left[E-\left(\frac{\mu \omega^{2}r^{2}}{2}+\frac{g}{r^{2}}\right)\right]R(r)=0~.
\end{align}
In our case we have a one-dimensional harmonic oscillator with inverse quadratic potential. Thus, we may impose the following choices: $N=1$ and $\ell=0$, in the above differential equation. For which, the above differential equation boils down to, 
\begin{align}\label{eq:alternate_equation}
R''(r)+\frac{2\mu}{\hbar^{2}}\left[E-\left(\frac{\mu \omega^{2}r^{2}}{2}+\frac{g}{r^{2}}\right)\right]R(r)=0~.
\end{align}
As in the Wheeler-deWitt equation presented in \ref{WdW_II}, the above equation has both $r^{2}$ and $(1/r^{2})$ term, and hence the parameters appearing in \ref{eq:alternate_equation} can be related to those appearing in \ref{WdW_II}, through the following identifications,
\begin{align}\label{identifications}
E=\frac{1}{2}~;\qquad 
\mu=1~;\qquad 
\omega =1~;\qquad 
g=\frac{\sigma^2}{8}~;\qquad
\frac{1}{\hbar}=i\alpha_n~.
\end{align}
For the moment being we will keep the analysis general and shall finally use the above identifications in order to map the scenario to the one of our interest. First of all, performing a change of variable, 
\begin{align}
R(r)=r^{\gamma}\exp \left(-\frac{\mu \omega r^{2}}{2\hbar}\right)F(r)~,
\end{align}
we obtain,
\begin{align}
R'(r)&=\gamma r^{\gamma-1}\exp \left(-\frac{\mu \omega r^{2}}{2\hbar}\right)F(r)+r^{\gamma}\exp \left(-\frac{\mu \omega r^{2}}{2\hbar}\right)F'(r)+r^{\gamma}\exp \left(-\frac{\mu \omega r^{2}}{2\hbar}\right)F(r)\left(-\frac{\mu \omega r}{\hbar}\right)~,
\\
R''(r)&=\gamma\left(\gamma-1\right)r^{\gamma-2}\exp \left(-\frac{\mu \omega r^{2}}{2\hbar}\right)F(r)+2\gamma r^{\gamma-1}\exp \left(-\frac{\mu \omega r^{2}}{2\hbar}\right)F'(r)
\nonumber
\\
&\hskip 1 cm +2\gamma r^{\gamma-1}\exp \left(-\frac{\mu \omega r^{2}}{2\hbar}\right)F(r)\left(-\frac{\mu \omega r}{\hbar}\right)
+r^{\gamma}\exp \left(-\frac{\mu \omega r^{2}}{2\hbar}\right)F''(r)
\nonumber
\\
&\hskip 1 cm +2r^{\gamma}\exp \left(-\frac{\mu \omega r^{2}}{2\hbar}\right)F'(r)\left(-\frac{\mu \omega r}{\hbar}\right)+r^{\gamma}\exp \left(-\frac{\mu \omega r^{2}}{2\hbar}\right)F(r)\left(-\frac{\mu \omega r}{\hbar}\right)^{2}
\nonumber
\\
&\hskip 1 cm +r^{\gamma}\exp \left(-\frac{\mu \omega r^{2}}{2\hbar}\right)F(r)\left(-\frac{\mu \omega }{\hbar}\right)~.
\end{align}
Such that, the substitution of the above expressions for $R''(r)$ and $R'(r)$ in \ref{eq:alternate_equation}, yields,
\begin{align}
&\gamma\left(\gamma-1\right)r^{\gamma-2}\exp \left(-\frac{\mu \omega r^{2}}{2\hbar}\right)F(r)+2\gamma r^{\gamma-1}\exp \left(-\frac{\mu \omega r^{2}}{2\hbar}\right)F'(r)
\nonumber
\\
&\hskip 1 cm +2\gamma r^{\gamma-1}\exp \left(-\frac{\mu \omega r^{2}}{2\hbar}\right)F(r)\left(-\frac{\mu \omega r}{\hbar}\right)
+r^{\gamma}\exp \left(-\frac{\mu \omega r^{2}}{2\hbar}\right)F''(r)
\nonumber
\\
&\hskip 1 cm +2r^{\gamma}\exp \left(-\frac{\mu \omega r^{2}}{2\hbar}\right)F'(r)\left(-\frac{\mu \omega r}{\hbar}\right)+r^{\gamma}\exp \left(-\frac{\mu \omega r^{2}}{2\hbar}\right)F(r)\left(-\frac{\mu \omega r}{\hbar}\right)^{2}
\nonumber
\\
&\hskip 1 cm +r^{\gamma}\exp \left(-\frac{\mu \omega r^{2}}{2\hbar}\right)F(r)\left(-\frac{\mu \omega }{\hbar}\right)+\frac{2\mu}{\hbar^{2}}\left[E-\left(\frac{\mu \omega^{2}r^{2}}{2}+\frac{g}{r^{2}}\right)\right]r^{\gamma}\exp \left(-\frac{\mu \omega r^{2}}{2\hbar}\right)F(r)=0~,
\end{align}
which simplifies to the following differential equation,
\begin{align}
F''(r)&+\left[\frac{2\gamma}{r}-\frac{2\mu \omega r}{\hbar} \right]F'(r)
\nonumber
\\
&\hskip 1 cm +\left[\frac{\gamma\left(\gamma-1\right)}{r^{2}}-\frac{2\gamma\mu \omega}{\hbar}-\frac{\mu \omega}{\hbar}+\frac{\mu^{2}\omega^{2}r^{2}}{\hbar^{2}}+\frac{2\mu E}{\hbar^{2}}-\frac{\mu^{2}\omega^{2}r^{2}}{\hbar^{2}}-\frac{2\mu g}{\hbar^{2}r^{2}} \right]F(r)=0~.
\end{align}
Further simplification yields the following differential equation,
\begin{align}
F''(r)&+\left[\frac{2\gamma}{r}-\frac{2\mu \omega r}{\hbar} \right]F'(r)+\left[-\frac{2\gamma\mu \omega}{\hbar}-\frac{\mu \omega}{\hbar}+\frac{2\mu E}{\hbar^{2}}\right]F(r)=0~,
\end{align}
where, we have fixed the constant $\gamma$, such that it satisfies the following algebraic relation: $\gamma(\gamma-1)=(2\mu g/\hbar^{2})$. If we make a further change of variable from $r$ to $\xi$, such that $\xi\equiv (\mu \omega r^{2}/\hbar)$, then it follows that, 
\begin{align}
\frac{dF}{dr}=\frac{dF}{d\xi}\left(\frac{2\mu \omega r}{\hbar}\right)~;
\qquad 
\frac{d^{2}F}{dr^{2}}=\left(\frac{2\mu \omega r}{\hbar}\right)\frac{d}{d\xi}\left[\left(\frac{2\mu \omega r}{\hbar}\right)\frac{dF}{d\xi}\right]
=\left(\frac{2\mu \omega r}{\hbar}\right)^{2}\frac{d^{2}F}{d\xi^{2}}+\left(\frac{2\mu \omega}{\hbar}\right)\frac{dF}{d\xi}~.
\end{align}
Thus the differential equation takes the form,
\begin{align}
\left(\frac{2\mu \omega r}{\hbar}\right)^{2}\frac{d^{2}F}{d\xi^{2}}+\left(\frac{2\mu \omega}{\hbar}\right)\frac{dF}{d\xi}+\left[\frac{2\alpha_{0}}{r}-\frac{2\mu \omega r}{\hbar} \right]\frac{dF}{d\xi}\left(\frac{2\mu \omega r}{\hbar}\right)
+\left[-\frac{2\alpha_{0}\mu \omega}{\hbar}-\frac{\mu \omega}{\hbar}+\frac{2\mu E}{\hbar^{2}}\right]F(r)=0~,
\end{align}
which can be re-expressed solely in terms of the variable $\xi$ as,
\begin{align}
\frac{4\mu \omega}{\hbar}\xi\frac{d^{2}F}{d\xi^{2}}+\left[\frac{4\alpha_{0}\mu \omega}{\hbar}-\frac{4\mu \omega \xi }{\hbar}+\frac{2\mu \omega}{\hbar} \right]\frac{dF}{d\xi}+\left[-\frac{2\alpha_{0}\mu \omega}{\hbar}-\frac{\mu \omega}{\hbar}+\frac{2\mu E}{\hbar^{2}}\right]F(r)=0~.
\end{align}
Finally, this yields,
\begin{align}
\xi\frac{d^{2}F}{d\xi^{2}}+\left[\left(\alpha_{0}+\frac{1}{2}\right)-\xi\right]\frac{dF}{d\xi}+\left[-\frac{\alpha_{0}}{2}-\frac{1}{4}+\frac{E}{2\hbar \omega}\right]F(r)=0~.
\end{align}
Using the identifications with the Wheeler-deWitt equation, as presented in \ref{identifications}, the above differential equation can be re-written as,
\begin{align}
\xi\frac{d^2F}{d\xi^2}+\left[\left(\gamma+\frac{1}{2}\right)-\xi\right]\frac{dF}{d\xi}-\left(\frac{1}{4}+\frac{\gamma}{2}-\frac{i\alpha_n}{4}\right)F(\xi)=0~.
\end{align}
Solution to the above equation is in terms of the confluent hypergeometric function, such that,
\begin{align}
F(\xi)=M\left(\frac{1}{4}+\frac{\gamma}{2}-\frac{i\alpha_n}{4},\gamma+\frac{1}{2},\xi\right)~.
\end{align}
Now, we have the following identities,
\begin{align}
\frac{1}{4}+\frac{\gamma}{2}-\frac{i\alpha_n}{4}&=\frac{1}{4}+\frac{1}{4}\pm\frac{1}{4}\sqrt{1-\alpha_n^2\sigma^2}-\frac{i\alpha_n}{4}=\frac{1}{2}+\mu_2-\mu_1~;
\\
\gamma+\frac{1}{2}&=\frac{1}{2}+\frac{1}{2}\pm\frac{1}{2}\sqrt{1-\alpha_n^2\sigma^2}=1+2\mu_2~,
\end{align}
where, we have defined, $\mu_2=\pm(1/4)\sqrt{1-\alpha_n^2\sigma^2}$ and $\mu_1=(i\alpha_n/4)$. Therefore, the solution for $R(r)$ takes the following form,
\begin{align}
R(\xi)&=r^{\gamma}\exp\left(-\frac{\xi}{2}\right)F(\xi)
\nonumber
\\
&=\xi^{\frac{1}{4}+\mu_2}\exp\left(-\frac{\xi}{2}\right)M\left(\frac{1}{2}+\mu_2-\mu_1,1+2\mu_2,\xi\right)~.
\end{align}
Using the following relation between the Whittaker function and the confluent Hypergeometric function,
\begin{align}
M_{\mu_1,\mu_2}(\xi)=\xi^{\frac{1}{2}+\mu_2}\exp\left(-\frac{\xi}{2}\right)M\left(\frac{1}{2}+\mu_2-\mu_1,1+2\mu_2,\xi\right)~,
\end{align}
we obtain,
\begin{align}
\psi_{\vec{\Pi}}(q)\propto (i\alpha_n q^2)^{-\frac{1}{4}}M_{\mu_1,\mu_2}(i\alpha_n q^2)~,
\end{align}
where, $R(r)$ has been identified with the Wheeler-deWitt wave function $\psi_{\vec{\Pi}}(q)$ and as evident the solution coincides with the one presented in \ref{section_WdW_solution}. Hence, the alternative approach, presented here is equivalent to the one used in the main text.
\section{Lorentzian path integral approach: Details of the computation}\label{App_My}

In this appendix we will solve for the Kernel of the harmonic oscillator, living within an inverse quadratic potential, which will be used extensively in the Lorentzian path integral description of the main text. For this purpose, from the mini-superspace action presented in \ref{action_ueff}, we obtain the following mini-superspace Lagrangian,  
\begin{align}
\mathcal{L}=-\frac{MV_{3}}{2\mathcal{N}}\dot{q}^{2}+\frac{V_{3}q^{2}}{2\mathcal{N}}\left(\dot{\theta}_{1}^{2}+\dot{\theta}_{2}^{2}\right)+\mathcal{N}V_{3}U_{\rm eff}~.
\end{align}
The gravitational Hamiltonian can be derived from the variational derivative of the above Lagrangian with respect to the Lapse function $N$, which reads in terms of the dynamical variable $q$ as,
\begin{align}
\mathcal{H}_{\rm grav}=\frac{\delta \mathcal{L}}{\delta \mathcal{N}}\Big\vert_{\mathcal{N}=1}=\frac{MV_{3}}{2}\dot{q}^{2}-\frac{V_{3}q^{2}}{2}\left(\dot{\theta}_{1}^{2}+\dot{\theta}_{2}^{2}\right)+V_{3}U_{\rm eff}~.
\end{align}
Similarly, the momenta conjugate to the dynamical variables $q$, $\theta_{1}$ and $\theta_{2}$, associated with the above Lagrangian, becomes, 
\begin{align}
p_{q}=-MV_{3}\dot{q}~;\qquad p_{1}=V_{3}q^{2}\dot{\theta}_{1}~;\quad p_{2}=V_{3}q^{2}\dot{\theta}_{2}~.
\end{align}
Thus, the gravitational Hamiltonian can also be written as,
\begin{align}
\mathcal{H}_{\rm grav}=\frac{p_{q}^{2}}{2MV_{3}}-\frac{p_{1}^{2}+p_{2}^{2}}{2V_{3}q^{2}}+V_{3}U_{\rm eff}~.
\end{align}
Introducing, $p_{q}=-i\hbar(\partial/\partial q)$, and similarly for the momenta conjugate to $\theta_{1}$ and $\theta_{2}$ respectively, we obtain the operator form of the gravitational Hamiltonian,
\begin{align}
\mathcal{H}_{\rm grav}=-\frac{\hbar^{2}}{2MV_{3}}\frac{\partial^{2}}{\partial q^{2}}+\frac{\hbar^{2}}{2V_{3}q^{2}}\left(\frac{\partial^{2}}{\partial \theta_{1}^{2}}+ \frac{\partial^{2}}{\partial \theta_{2}^{2}}\right)+V_{3}U_{\rm eff}~.
\end{align}
On the other hand, the Hamiltonian $\mathcal{H}_{\bm{Q}}$ associated with the degrees of freedom $\bm{Q}\equiv (q,\theta_{1},\theta_{2})$, takes the following form,
\begin{align}
\mathcal{H}_{\bm{Q}}&=p_{q}\dot{q}+p_{1}\dot{\theta}_{1}+p_{2}\dot{\theta_{2}}-\mathcal{L}
\nonumber
\\
&=-MV_{3}\dot{q}^{2}+V_{3}q^{2}\left(\dot{\theta}_{1}^{2}+\dot{\theta}_{2}^{2}\right)+\frac{MV_{3}}{2}\dot{q}^{2}-\frac{V_{3}q^{2}}{2}\left(\dot{\theta}_{1}^{2}+\dot{\theta}_{2}^{2}\right)-V_{3}U_{\rm eff}
\nonumber
\\
&=-\frac{MV_{3}}{2}\dot{q}^{2}+\frac{V_{3}q^{2}}{2}\left(\dot{\theta}_{1}^{2}+\dot{\theta}_{2}^{2}\right)-V_{3}U_{\rm eff}=-\mathcal{H}_{\rm grav}~.
\end{align}
The above Hamiltonian $\mathcal{H}_{\bm{Q}}$ can also be transformed to an operator form, following the same strategy as that of the gravitational Hamiltonian. Given this, one introduces the propagator $\mathcal{G}_{\rm full}(\bm{Q},\bm{Q'};t,t')$ (in the gauge $\mathcal{N}=1$), such that it satisfies the following differential equation,  
\begin{align}\label{full_kernel}
i\hbar\partial_{t}\mathcal{G}_{\rm full}=\mathcal{H}_{\bm{Q}}\mathcal{G}_{\rm full}~.
\end{align}
Since the coordinates $\theta_{1}$ and $\theta_{2}$ are cyclic, it follows that we can decompose the propagator, such that 
\begin{align}
\mathcal{G}_{\rm full}(\bm{Q},\bm{Q}';t,t')=\int d^{2}\vec{\Pi}~\mathcal{G}_{\vec{\Pi}}(q,q';t,t')\exp\Big[\frac{iV_{3}}{\hbar}\vec{\Pi}\cdot\left(\vec{\theta}-\vec{\theta}'\right)\Big]~,
\end{align}
such that from \ref{full_kernel}, it follows that the reduced propagator $\mathcal{G}_{\vec{\Pi}}(q,q';t,t')$ satisfies the following differential equation,
\begin{align}
i\hbar\partial_{t}\mathcal{G}_{\vec{\Pi}}=\left[\frac{\hbar^{2}}{2MV_{3}}\frac{\partial^{2}}{\partial q^{2}}+\frac{V_{3}|\vec{\Pi}|^{2}}{2q^{2}}-V_{3}U_{\rm eff}\right]\mathcal{G}_{\vec{\Pi}}~.
\end{align}
The above differential equation can be re-written as,
\begin{align}
i\hbar\partial_{t}\mathcal{G}_{\vec{\Pi}}=\left[\frac{\hbar^{2}}{2MV_{3}}\frac{\partial^{2}}{\partial q^{2}}-V_{3}\rho_{0}\left(1-q^{2}-\frac{\sigma^{2}}{4q^{2}}\right)\right]\mathcal{G}_{\vec{\Pi}}~,
\end{align}
where we have used the following definition, $\rho_{\theta}=(|\vec{\Pi}|^{2}/2)$ and $(\sigma^{2}/4)=(\rho_{\theta}/\rho_{0})$. Further, using, $h_{n}^{2}=(2\rho_{0}/M)$, from \ref{def_h}, we obtain,
\begin{align}
i\hbar\partial_{t}\mathcal{G}_{\vec{\Pi}}=\left[\frac{\hbar^{2}}{2MV_{3}}\frac{\partial^{2}}{\partial q^{2}}-\frac{MV_{3}h_{n}^{2}}{2}\left(1-q^{2}-\frac{\sigma^{2}}{4q^{2}}\right)\right]\mathcal{G}_{\vec{\Pi}}~.
\end{align}
Thus we have derived the relevant equation satisfied by the reduced propagator $\mathcal{G}_{\vec{\Pi}}$. Further, we may decompose the propagator as, $\mathcal{G}_{\vec{\Pi}}=\exp(iMV_{3}h_{n}^{2}t/2\hbar)\widetilde{\mathcal{G}}_{\vec{\Pi}}$, such that $\widetilde{\mathcal{G}}_{\vec{\Pi}}$ satisfies the following differential equation,
\begin{align}\label{new1}
i\hbar\partial_{t}\widetilde{\mathcal{G}}_{\vec{\Pi}}=\left[\frac{\hbar^{2}}{2MV_{3}}\frac{\partial}{\partial q^{2}}+\frac{1}{2}MV_{3}h_{n}^{2}q^{2}+\frac{MV_{3}h_{n}^{2}\sigma^{2}}{8q^{2}}\right]\widetilde{\mathcal{G}}_{\vec{\Pi}}~.
\end{align}
We will try to find out a solution for $\widetilde{\mathcal{G}}_{\vec{\Pi}}$ in what follows. 

For this purpose, we would like to point out that in \cite{ChengandChan} the propagator associated with the following Schr\"{o}dinger equation,
\begin{align}\label{cheng1}
i\hbar\partial_{t}\mathcal{G}=\left[-\frac{\hbar^{2}}{2M}\frac{\partial}{\partial q^{2}}+\frac{1}{2}M\omega^{2}q^{2}+\frac{g}{q^{2}}\right]\mathcal{G}~,
\end{align}
was obtained as,
\begin{align}\label{cheng2}
\mathcal{G}(q_{f},t_{f};q_{i},t_{i})=\left(\frac{-iM\sqrt{q_{f}q_{i}}}{\hbar a_{f}} \right)\exp \left[\frac{i}{2\hbar a_{f}}\left(Mb_{f}q_{i}^{2}+M\dot{a}_{f}q_{f}^{2} \right) \right]I_{\nu}\left(-\frac{iMq_{i}q_{f}}{\hbar a_{f}}\right)~,
\end{align}
where, $\nu=(1/2)\sqrt{1+(8Mg)/\hbar^{2}}$, $I_{\nu}$ is the modified Bessel function of order $\nu$ and the functions $a(t)$ and $b(t)$ satisfies the following differential equations, with certain initial conditions,
\begin{align}
\ddot{a}+\omega^{2}a=0~;\qquad a_{i}=0~,\dot{a}_{i}=1~.
\label{cheng_3a}
\\
\ddot{b}+\omega^{2}b=0~;\qquad b_{i}=1~,\dot{b}_{i}=0~.
\label{cheng_3b}
\end{align}
where, `dot' denotes derivative with respect to the time coordinate $t$ and subscript $i$ denotes the value of the function evaluated at $t=t_{i}$. A comparison of \ref{new1} with \ref{cheng1}, reveals the following correspondence,
\begin{align}\label{new2}
M\leftrightarrow -MV_{3}~;\qquad
\omega^{2}\leftrightarrow -h_{n}^{2}~;\qquad
g\leftrightarrow \frac{MV_{3}h_{n}^{2}\sigma^{2}}{8}~.
\end{align}
Thus the order of the modified Bessel function becomes,
\begin{align}
\nu=\frac{1}{2}\sqrt{1-\frac{M^{2}V^{2}_{3}h_{n}^{2}\sigma^{2}}{\hbar^{2}}}=\frac{1}{2}\sqrt{1-\frac{\sigma^{2}}{\sigma_{\rm c}^{2}}}
\end{align}
where, we have used the result that, $\sigma_{c}^{2}=(1/\alpha_{n}^{2})=(\hbar^{2}/M^{2}V^{2}_{3}h_{n}^{2})$. The equivalent equations to \ref{cheng_3a} and \ref{cheng_3b}, in the context of our interest becomes,
\begin{align}
\ddot{a}-h_{n}^{2}a=0~;\qquad a_{i}=0~,\dot{a}_{i}=1~.
\label{new_3a}
\\
\ddot{b}-h_{n}^{2}b=0~;\qquad b_{i}=1~,\dot{b}_{i}=0~.
\label{new_3b}
\end{align}
with the following solutions,
\begin{align}
a(t)=\frac{1}{h_{n}}\sinh \left[h_{n}(t-t_{i}) \right]~;\qquad
b(t)=\cosh \left[h_{n}(t-t_{i}) \right]~.
\end{align}
However it is customary to introduce a new time coordinate $T=\mathcal{N}t$, in terms of which, from \ref{cheng2}, the propagator $\widetilde{\mathcal{G}}_{\vec{\Pi}}$ becomes,
\begin{align}\label{newkernel1}
\widetilde{\mathcal{G}}_{\vec{\Pi}}(q_{f},T_{f};q_{i},T_{i})&=\left(\frac{iMV_{3}h_{n}\sqrt{q_{f}q_{i}}}{\hbar \sinh \left[h_{n}(T_{f}-T_{i}) \right]} \right)
~I_{\nu}\left(\frac{iMV_{3}h_{n}q_{i}q_{f}}{\hbar \sinh \left[h_{n}(T_{f}-T_{i}) \right]}\right)
\nonumber
\\
&\hskip 0.5 cm \times 
\exp \left[\frac{-ih_{n}}{2\hbar \sinh \left[h_{n}(T_{f}-T_{i}) \right]}\left(MV_{3}\cosh \left[h_{n}(T_{f}-T_{i}) \right]q_{i}^{2}+MV_{3}\cosh \left[h_{n}(T_{f}-T_{i}) \right]q_{f}^{2} \right) \right]
\nonumber
\\
&=\left(\frac{iMV_{3}h_{n}\sqrt{q_{f}q_{i}}}{\hbar \sinh \left[h_{n}(T_{f}-T_{i}) \right]} \right)\exp \left[-\frac{ih_{n}MV_{3}}{2\hbar}\coth \left[h_{n}(T_{f}-T_{i}) \right] \left(q_{i}^{2}+q_{f}^{2} \right) \right]
\nonumber
\\
&\hskip 2 cm \times I_{\nu}\left(\frac{iMV_{3}h_{n}q_{i}q_{f}}{\hbar \sinh \left[h_{n}(T_{f}-T_{i}) \right]}\right)
\end{align}
Introducing the factor $\exp(iMV_{3}h_{n}^{2}t/2\hbar)$, the final expression for the propagator reads,
\begin{align}\label{newkernel2}
\mathcal{G}_{\vec{\Pi}}(q_{f},T_{f};q_{i},T_{i})&=\exp\left[\frac{iMV_{3}h_{n}^{2}\left(T_{f}-T_{i}\right)}{2\hbar}\right]\widetilde{\mathcal{K}}_{\vec{\Pi}}(q_{f},T_{f};q_{i},T_{i})
\nonumber
\\
&=\left(\frac{iMV_{3}h_{n}\sqrt{q_{f}q_{i}}}{\hbar \sinh \left[h_{n}(T_{f}-T_{i}) \right]} \right)\exp \left[-\frac{ih_{n}MV_{3}}{2\hbar}\coth \left[h_{n}(T_{f}-T_{i}) \right] \left(q_{i}^{2}+q_{f}^{2} \right)+\frac{iMV_{3}h_{n}^{2}\left(T_{f}-T_{i}\right)}{2\hbar} \right]
\nonumber
\\
&\hskip 2 cm \times I_{\nu}\left(\frac{iMV_{3}h_{n}q_{i}q_{f}}{\hbar \sinh \left[h_{n}(T_{f}-T_{i}) \right]}\right)
\end{align}
Note that, the above propagator depends only on the combination $(T_{f}-T_{i})$. Further introducing the constant, $\alpha_{n}\equiv MV_{3}h_{n}/\hbar$, the expression for the propagator can be further simplified, yielding \ref{gen_kernel} of the main text. 

In what follows, we will also provide the derivation of the wave function, which in the semi-classical limit yields the solution of the Wheeler-deWitt equation. As emphasized in the main text, this requires a seed wave function $\phi_{0}(q)$, which is chosen to be the one presented in \ref{seed}. This yields the wave function $\phi_{\vec{\Pi}}(q)$ as,
\begin{align}
\phi_{\vec{\Pi}}(q)&=\int_{-\infty}^{\infty}dT\underbrace{\int_{0}^{\infty}\mathcal{G}_{\vec{\Pi}}(q,q';T-T_0)\phi_{0}(q')dq'}_{\mathcal{K}(q;T,T_{0})}~.
\end{align}
Substituting for $\mathcal{G}_{\vec{\Pi}}$ from \ref{gen_kernel} and using \ref{seed}, the function $\mathcal{K}(q;T,T_{0})$ takes the following form,  
\begin{align}
\mathcal{K}(q;T,T_{0})&=\left(\frac{i\alpha_{n}\sqrt{qq_{0}}}{\sinh \left[h_{n}(T-T_{0}) \right]} \right)\exp \left[-\frac{i\alpha_{n}}{2}\coth \left[h_{n}(T-T_{0}) \right]q^{2}+\frac{i\alpha_{n}h_{n}\left(T-T_{0}\right)}{2} \right]
\nonumber
\\
& \hskip 1 cm  \int_{0}^{\infty} dq' q'\exp \left[-\frac{i\alpha_{n}}{2}\left\{1+\coth \left[h_{n}(T-T_{0}) \right]\right\}q'^{2}\right]I_{\nu}\left(\frac{i\alpha_{n}qq'}{\sinh \left[h_{n}(T-T_{0}) \right]}\right)I_{\nu}\left(-i\alpha_{n}q'q_{0}e^{-h_{n}T_{0}}\right)
\nonumber
\\
\nonumber
\\
&=\left(\frac{i\alpha_{n}\sqrt{qq_{0}}}{\sinh \left[h_{n}(T-T_{0}) \right]} \right)\exp \left[-\frac{i\alpha_{n}}{2}\coth \left[h_{n}(T-T_{0}) \right]q^{2}+\frac{i\alpha_{n}h_{n}\left(T-T_{0}\right)}{2} \right]
\nonumber
\\
&\hskip 1 cm \left(\frac{1}{i\alpha_{n}\left\{1+\coth \left[h_{n}(T-T_{0}) \right]\right\}}\right)\exp\left[\frac{-\frac{\alpha_{n}^{2}q^{2}}{\sinh^{2} \left[h_{n}(T-T_{0}) \right]}-\alpha_{n}^{2}q_{0}^{2}e^{-2h_{n}T_{0}}}{2i\alpha_{n}\left\{1+\coth \left[h_{n}(T-T_{0}) \right]\right\}} \right]
\nonumber
\\
&\hskip 1 cm \times I_{\nu}\left(\frac{\alpha_{n}^{2}qq_{0}e^{-h_{n}T_{0}}}{\sinh \left[h_{n}(T-T_{0}) \right]i\alpha_{n}\left\{1+\coth \left[h_{n}(T-T_{0}) \right]\right\}} \right)
\nonumber
\\
\nonumber
\\
&=\sqrt{qq_{0}}e^{-h_{n}(T-T_{0})}\exp \left[\frac{i\alpha_{n}h_{n}\left(T-T_{0}\right)}{2} \right]I_{\nu}\left(-i\alpha_{n}qq_{0}e^{-h_{n}T} \right)
\nonumber
\\
&\hskip 1 cm \times \exp \left[-\frac{i\alpha_{n}}{2}q^{2}\right]\exp\left[-\frac{i\alpha_{n}q_{0}^{2}(e^{-2h_{n}T}-e^{-2h_{n}T_{0}})}{4}\right]~.
\end{align}
This is the result we have used in order to arrive at \ref{seed_kernel} in the main text. We have also used the following results in order to arrive at the above expression for $\mathcal{K}(q;T,T_{0})$,
\begin{align}
\frac{1}{\sinh^{2}x(1+\coth x)}-\coth x&=\frac{e^{-x}}{\sinh x}-\coth x=\frac{e^{-x}-\cosh x}{\sinh x}=\frac{e^{-x}-e^{x}}{2\sinh x}=-1~,
\\
\frac{1}{(1+\coth x)}&=\frac{\sinh x}{\sinh x+\cosh x}=e^{-x}\sinh x=\frac{1-e^{-2x}}{2}~,
\\
\int_{0}^{\infty}dq~q~e^{-p^{2}q^{2}}I_{\nu}(aq)&I_{\nu}(bq)=\frac{1}{2p^{2}}\exp\left[\frac{a^{2}+b^{2}}{4p^{2}}\right]I_{\nu}\left(\frac{ab}{2p^{2}}\right)~.
\end{align}

\bibliography{Shear_Bounce_Quantum}

\providecommand{\href}[2]{#2}\begingroup\raggedright\begin{thebibliography}{10}

\bibitem{Guth:1980zm}
A.~H. Guth, ``{The Inflationary Universe: A Possible Solution to the Horizon
  and Flatness Problems},''
  \href{http://dx.doi.org/10.1103/PhysRevD.23.347}{{\em Phys. Rev. D}
  {\bfseries 23} (1981) 347--356}.

\bibitem{Sato:1980yn}
K.~Sato, ``{First Order Phase Transition of a Vacuum and Expansion of the
  Universe},'' {\em Mon. Not. Roy. Astron. Soc.} {\bfseries 195} (1981)
  467--479.

\bibitem{Linde:1981mu}
A.~D. Linde, ``{A New Inflationary Universe Scenario: A Possible Solution of
  the Horizon, Flatness, Homogeneity, Isotropy and Primordial Monopole
  Problems},'' \href{http://dx.doi.org/10.1016/0370-2693(82)91219-9}{{\em Phys.
  Lett. B} {\bfseries 108} (1982) 389--393}.

\bibitem{Albrecht:1982wi}
A.~Albrecht and P.~J. Steinhardt, ``{Cosmology for Grand Unified Theories with
  Radiatively Induced Symmetry Breaking},''
  \href{http://dx.doi.org/10.1103/PhysRevLett.48.1220}{{\em Phys. Rev. Lett.}
  {\bfseries 48} (1982) 1220--1223}.

\bibitem{Starobinsky:1980te}
A.~A. Starobinsky, ``{A New Type of Isotropic Cosmological Models Without
  Singularity},'' \href{http://dx.doi.org/10.1016/0370-2693(80)90670-X}{{\em
  Phys. Lett. B} {\bfseries 91} (1980) 99--102}.

\bibitem{Borde:1993xh}
A.~Borde and A.~Vilenkin, ``{Eternal inflation and the initial singularity},''
  \href{http://dx.doi.org/10.1103/PhysRevLett.72.3305}{{\em Phys. Rev. Lett.}
  {\bfseries 72} (1994) 3305--3309},
  \href{http://arxiv.org/abs/gr-qc/9312022}{{\ttfamily arXiv:gr-qc/9312022}}.

\bibitem{Martin:2000xs}
J.~Martin and R.~H. Brandenberger, ``{The TransPlanckian problem of
  inflationary cosmology},''
  \href{http://dx.doi.org/10.1103/PhysRevD.63.123501}{{\em Phys. Rev. D}
  {\bfseries 63} (2001) 123501},
  \href{http://arxiv.org/abs/hep-th/0005209}{{\ttfamily arXiv:hep-th/0005209}}.

\bibitem{Brandenberger_2013}
R.~H. Brandenberger and J.~Martin, ``Trans-planckian issues for inflationary
  cosmology,'' \href{http://dx.doi.org/10.1088/0264-9381/30/11/113001}{{\em
  Classical and Quantum Gravity} {\bfseries 30} no.~11, (Apr, 2013) 113001}.
  \url{http://dx.doi.org/10.1088/0264-9381/30/11/113001}.

\bibitem{Novello:2008ra}
M.~Novello and S.~E.~P. Bergliaffa, ``{Bouncing Cosmologies},''
  \href{http://dx.doi.org/10.1016/j.physrep.2008.04.006}{{\em Phys. Rept.}
  {\bfseries 463} (2008) 127--213},
  \href{http://arxiv.org/abs/0802.1634}{{\ttfamily arXiv:0802.1634
  [astro-ph]}}.

\bibitem{Nandi:2019xag}
D.~Nandi and L.~Sriramkumar, ``{Can a nonminimal coupling restore the
  consistency condition in bouncing universes?},''
  \href{http://dx.doi.org/10.1103/PhysRevD.101.043506}{{\em Phys. Rev. D}
  {\bfseries 101} no.~4, (2020) 043506},
  \href{http://arxiv.org/abs/1904.13254}{{\ttfamily arXiv:1904.13254 [gr-qc]}}.

\bibitem{Raveendran:2018yyh}
R.~N. Raveendran and L.~Sriramkumar, ``{Primordial features from ekpyrotic
  bounces},'' \href{http://dx.doi.org/10.1103/PhysRevD.99.043527}{{\em Phys.
  Rev. D} {\bfseries 99} no.~4, (2019) 043527},
  \href{http://arxiv.org/abs/1809.03229}{{\ttfamily arXiv:1809.03229
  [astro-ph.CO]}}.

\bibitem{Lilley:2015ksa}
M.~Lilley and P.~Peter, ``{Bouncing alternatives to inflation},''
  \href{http://dx.doi.org/10.1016/j.crhy.2015.08.009}{{\em Comptes Rendus
  Physique} {\bfseries 16} (2015) 1038--1047},
  \href{http://arxiv.org/abs/1503.06578}{{\ttfamily arXiv:1503.06578
  [astro-ph.CO]}}.

\bibitem{Ijjas:2015hcc}
A.~Ijjas and P.~J. Steinhardt, ``{Implications of Planck2015 for inflationary,
  ekpyrotic and anamorphic bouncing cosmologies},''
  \href{http://dx.doi.org/10.1088/0264-9381/33/4/044001}{{\em Class. Quant.
  Grav.} {\bfseries 33} no.~4, (2016) 044001},
  \href{http://arxiv.org/abs/1512.09010}{{\ttfamily arXiv:1512.09010
  [astro-ph.CO]}}.

\bibitem{Raveendran:2017vfx}
R.~N. Raveendran, D.~Chowdhury, and L.~Sriramkumar, ``{Viable tensor-to-scalar
  ratio in a symmetric matter bounce},''
  \href{http://dx.doi.org/10.1088/1475-7516/2018/01/030}{{\em JCAP} {\bfseries
  01} (2018) 030}, \href{http://arxiv.org/abs/1703.10061}{{\ttfamily
  arXiv:1703.10061 [gr-qc]}}.

\bibitem{Chowdhury:2015cma}
D.~Chowdhury, V.~Sreenath, and L.~Sriramkumar, ``{The tensor bi-spectrum in a
  matter bounce},'' \href{http://dx.doi.org/10.1088/1475-7516/2015/11/002}{{\em
  JCAP} {\bfseries 11} (2015) 002},
  \href{http://arxiv.org/abs/1506.06475}{{\ttfamily arXiv:1506.06475
  [astro-ph.CO]}}.

\bibitem{Li:2016xjb}
Y.-B. Li, J.~Quintin, D.-G. Wang, and Y.-F. Cai, ``{Matter bounce cosmology
  with a generalized single field: non-Gaussianity and an extended no-go
  theorem},'' \href{http://dx.doi.org/10.1088/1475-7516/2017/03/031}{{\em JCAP}
  {\bfseries 03} (2017) 031}, \href{http://arxiv.org/abs/1612.02036}{{\ttfamily
  arXiv:1612.02036 [hep-th]}}.

\bibitem{Finelli:2001sr}
F.~Finelli and R.~Brandenberger, ``{On the generation of a scale invariant
  spectrum of adiabatic fluctuations in cosmological models with a contracting
  phase},'' \href{http://dx.doi.org/10.1103/PhysRevD.65.103522}{{\em Phys. Rev.
  D} {\bfseries 65} (2002) 103522},
  \href{http://arxiv.org/abs/hep-th/0112249}{{\ttfamily arXiv:hep-th/0112249}}.

\bibitem{Brandenberger:2012zb}
R.~H. Brandenberger, ``{The Matter Bounce Alternative to Inflationary
  Cosmology},'' \href{http://arxiv.org/abs/1206.4196}{{\ttfamily
  arXiv:1206.4196 [astro-ph.CO]}}.

\bibitem{Brandenberger:2016vhg}
R.~Brandenberger and P.~Peter, ``{Bouncing Cosmologies: Progress and
  Problems},'' \href{http://dx.doi.org/10.1007/s10701-016-0057-0}{{\em Found.
  Phys.} {\bfseries 47} no.~6, (2017) 797--850},
  \href{http://arxiv.org/abs/1603.05834}{{\ttfamily arXiv:1603.05834
  [hep-th]}}.

\bibitem{Pereira:2007yy}
T.~S. Pereira, C.~Pitrou, and J.-P. Uzan, ``{Theory of cosmological
  perturbations in an anisotropic universe},''
  \href{http://dx.doi.org/10.1088/1475-7516/2007/09/006}{{\em JCAP} {\bfseries
  09} (2007) 006}, \href{http://arxiv.org/abs/0707.0736}{{\ttfamily
  arXiv:0707.0736 [astro-ph]}}.

\bibitem{Grain:2020wro}
J.~Grain and V.~Vennin, ``{Unavoidable shear from quantum fluctuations in
  contracting cosmologies},''
  \href{http://dx.doi.org/10.1140/epjc/s10052-021-08932-0}{{\em Eur. Phys. J.
  C} {\bfseries 81} no.~2, (2021) 132},
  \href{http://arxiv.org/abs/2005.04222}{{\ttfamily arXiv:2005.04222
  [astro-ph.CO]}}.

\bibitem{Belinsky:1970ew}
V.~A. Belinsky, I.~M. Khalatnikov, and E.~M. Lifshitz, ``{Oscillatory approach
  to a singular point in the relativistic cosmology},''
  \href{http://dx.doi.org/10.1080/00018737000101171}{{\em Adv. Phys.}
  {\bfseries 19} (1970) 525--573}.

\bibitem{Agullo:2020iqv}
I.~Agullo, J.~Olmedo, and V.~Sreenath, ``{Observational consequences of Bianchi
  I spacetimes in loop quantum cosmology},''
  \href{http://dx.doi.org/10.1103/PhysRevD.102.043523}{{\em Phys. Rev. D}
  {\bfseries 102} no.~4, (2020) 043523},
  \href{http://arxiv.org/abs/2006.01883}{{\ttfamily arXiv:2006.01883 [gr-qc]}}.

\bibitem{Rajeev:2021lqk}
K.~Rajeev, V.~Mondal, and S.~Chakraborty, ``{No-boundary wave function,
  Wheeler-DeWitt equation, and path integral analysis of the bouncing quantum
  cosmology},'' \href{http://dx.doi.org/10.1103/PhysRevD.103.106008}{{\em Phys.
  Rev. D} {\bfseries 103} no.~10, (2021) 106008},
  \href{http://arxiv.org/abs/2101.02848}{{\ttfamily arXiv:2101.02848 [gr-qc]}}.

\bibitem{Battefeld:2014uga}
D.~Battefeld and P.~Peter, ``{A Critical Review of Classical Bouncing
  Cosmologies},'' \href{http://dx.doi.org/10.1016/j.physrep.2014.12.004}{{\em
  Phys. Rept.} {\bfseries 571} (2015) 1--66},
  \href{http://arxiv.org/abs/1406.2790}{{\ttfamily arXiv:1406.2790
  [astro-ph.CO]}}.

\bibitem{Hawking:1969sw}
S.~W. Hawking and R.~Penrose, ``{The Singularities of gravitational collapse
  and cosmology},'' \href{http://dx.doi.org/10.1098/rspa.1970.0021}{{\em Proc.
  Roy. Soc. Lond. A} {\bfseries 314} (1970) 529--548}.

\bibitem{Brandenberger:2009yt}
R.~Brandenberger, ``{Matter Bounce in Horava-Lifshitz Cosmology},''
  \href{http://dx.doi.org/10.1103/PhysRevD.80.043516}{{\em Phys. Rev. D}
  {\bfseries 80} (2009) 043516},
  \href{http://arxiv.org/abs/0904.2835}{{\ttfamily arXiv:0904.2835 [hep-th]}}.

\bibitem{Bamba:2013fha}
K.~Bamba, A.~N. Makarenko, A.~N. Myagky, S.~Nojiri, and S.~D. Odintsov,
  ``{Bounce cosmology from $F(R)$ gravity and $F(R)$ bigravity},''
  \href{http://dx.doi.org/10.1088/1475-7516/2014/01/008}{{\em JCAP} {\bfseries
  01} (2014) 008}, \href{http://arxiv.org/abs/1309.3748}{{\ttfamily
  arXiv:1309.3748 [hep-th]}}.

\bibitem{Desai:2015haa}
S.~Desai and N.~J. Pop\l{}awski, ``{Non-parametric reconstruction of an
  inflaton potential from
  Einstein\textendash{}Cartan\textendash{}Sciama\textendash{}Kibble gravity
  with particle production},''
  \href{http://dx.doi.org/10.1016/j.physletb.2016.02.014}{{\em Phys. Lett. B}
  {\bfseries 755} (2016) 183--189},
  \href{http://arxiv.org/abs/1510.08834}{{\ttfamily arXiv:1510.08834
  [astro-ph.CO]}}.

\bibitem{ArkaniHamed:2003uy}
N.~Arkani-Hamed, H.-C. Cheng, M.~A. Luty, and S.~Mukohyama, ``{Ghost
  condensation and a consistent infrared modification of gravity},''
  \href{http://dx.doi.org/10.1088/1126-6708/2004/05/074}{{\em JHEP} {\bfseries
  05} (2004) 074}, \href{http://arxiv.org/abs/hep-th/0312099}{{\ttfamily
  arXiv:hep-th/0312099}}.

\bibitem{Cai:2007qw}
Y.-F. Cai, T.~Qiu, Y.-S. Piao, M.~Li, and X.~Zhang, ``{Bouncing universe with
  quintom matter},''
  \href{http://dx.doi.org/10.1088/1126-6708/2007/10/071}{{\em JHEP} {\bfseries
  10} (2007) 071}, \href{http://arxiv.org/abs/0704.1090}{{\ttfamily
  arXiv:0704.1090 [gr-qc]}}.

\bibitem{Cai:2008qw}
Y.-F. Cai, T.-t. Qiu, R.~Brandenberger, and X.-m. Zhang, ``{A Nonsingular
  Cosmology with a Scale-Invariant Spectrum of Cosmological Perturbations from
  Lee-Wick Theory},'' \href{http://dx.doi.org/10.1103/PhysRevD.80.023511}{{\em
  Phys. Rev. D} {\bfseries 80} (2009) 023511},
  \href{http://arxiv.org/abs/0810.4677}{{\ttfamily arXiv:0810.4677 [hep-th]}}.

\bibitem{Raveendran:2018why}
R.~N. Raveendran and L.~Sriramkumar, ``{Viable scalar spectral tilt and
  tensor-to-scalar ratio in near-matter bounces},''
  \href{http://dx.doi.org/10.1103/PhysRevD.100.083523}{{\em Phys. Rev. D}
  {\bfseries 100} no.~8, (2019) 083523},
  \href{http://arxiv.org/abs/1812.06803}{{\ttfamily arXiv:1812.06803
  [astro-ph.CO]}}.

\bibitem{Bonanno:2017gji}
A.~Bonanno, G.~Gionti, S.~J., and A.~Platania, ``{Bouncing and emergent
  cosmologies from Arnowitt\textendash{}Deser\textendash{}Misner RG flows},''
  \href{http://dx.doi.org/10.1088/1361-6382/aaa535}{{\em Class. Quant. Grav.}
  {\bfseries 35} no.~6, (2018) 065004},
  \href{http://arxiv.org/abs/1710.06317}{{\ttfamily arXiv:1710.06317 [gr-qc]}}.

\bibitem{Platania:2020lqb}
A.~Platania, ``{From renormalization group flows to cosmology},''
  \href{http://dx.doi.org/10.3389/fphy.2020.00188}{{\em Front. in Phys.}
  {\bfseries 8} (2020) 188}, \href{http://arxiv.org/abs/2003.13656}{{\ttfamily
  arXiv:2003.13656 [gr-qc]}}.

\bibitem{Platania:2019qvo}
A.~Platania, ``{The inflationary mechanism in Asymptotically Safe Gravity},''
  \href{http://dx.doi.org/10.3390/universe5080189}{{\em Universe} {\bfseries 5}
  no.~8, (2019) 189}, \href{http://arxiv.org/abs/1908.03897}{{\ttfamily
  arXiv:1908.03897 [gr-qc]}}.

\bibitem{Bamba:2014zoa}
K.~Bamba, A.~N. Makarenko, A.~N. Myagky, and S.~D. Odintsov, ``{Bounce universe
  from string-inspired Gauss-Bonnet gravity},''
  \href{http://dx.doi.org/10.1088/1475-7516/2015/04/001}{{\em JCAP} {\bfseries
  04} (2015) 001}, \href{http://arxiv.org/abs/1411.3852}{{\ttfamily
  arXiv:1411.3852 [hep-th]}}.

\bibitem{Basile:2021amb}
I.~Basile and A.~Platania, ``{Cosmological $\alpha'$-corrections from the
  functional renormalization group},''
  \href{http://arxiv.org/abs/2101.02226}{{\ttfamily arXiv:2101.02226
  [hep-th]}}.

\bibitem{Brandenberger:1988aj}
R.~H. Brandenberger and C.~Vafa, ``{Superstrings in the Early Universe},''
  \href{http://dx.doi.org/10.1016/0550-3213(89)90037-0}{{\em Nucl. Phys. B}
  {\bfseries 316} (1989) 391--410}.

\bibitem{Haro:2015oqa}
J.~Haro, A.~N. Makarenko, A.~N. Myagky, S.~D. Odintsov, and V.~K. Oikonomou,
  ``{Bouncing loop quantum cosmology in Gauss-Bonnet gravity},''
  \href{http://dx.doi.org/10.1103/PhysRevD.92.124026}{{\em Phys. Rev. D}
  {\bfseries 92} no.~12, (2015) 124026},
  \href{http://arxiv.org/abs/1506.08273}{{\ttfamily arXiv:1506.08273 [gr-qc]}}.

\bibitem{Ashtekar:2008ay}
A.~Ashtekar, ``{Singularity Resolution in Loop Quantum Cosmology: A Brief
  Overview},'' \href{http://dx.doi.org/10.1088/1742-6596/189/1/012003}{{\em J.
  Phys. Conf. Ser.} {\bfseries 189} (2009) 012003},
  \href{http://arxiv.org/abs/0812.4703}{{\ttfamily arXiv:0812.4703 [gr-qc]}}.

\bibitem{WilsonEwing:2012pu}
E.~Wilson-Ewing, ``{The Matter Bounce Scenario in Loop Quantum Cosmology},''
  \href{http://dx.doi.org/10.1088/1475-7516/2013/03/026}{{\em JCAP} {\bfseries
  03} (2013) 026}, \href{http://arxiv.org/abs/1211.6269}{{\ttfamily
  arXiv:1211.6269 [gr-qc]}}.

\bibitem{Cai:2014zga}
Y.-F. Cai and E.~Wilson-Ewing, ``{Non-singular bounce scenarios in loop quantum
  cosmology and the effective field description},''
  \href{http://dx.doi.org/10.1088/1475-7516/2014/03/026}{{\em JCAP} {\bfseries
  03} (2014) 026}, \href{http://arxiv.org/abs/1402.3009}{{\ttfamily
  arXiv:1402.3009 [gr-qc]}}.

\bibitem{York:1972sj}
J.~W. York, Jr., ``{Role of conformal three geometry in the dynamics of
  gravitation},'' \href{http://dx.doi.org/10.1103/PhysRevLett.28.1082}{{\em
  Phys. Rev. Lett.} {\bfseries 28} (1972) 1082--1085}.

\bibitem{Gibbons:1976ue}
G.~W. Gibbons and S.~W. Hawking, ``{Action Integrals and Partition Functions in
  Quantum Gravity},'' \href{http://dx.doi.org/10.1103/PhysRevD.15.2752}{{\em
  Phys. Rev. D} {\bfseries 15} (1977) 2752--2756}.

\bibitem{Chakraborty:2019doh}
S.~Chakraborty and T.~Padmanabhan, ``{Boundary term in the gravitational action
  is the heat content of the null surfaces},''
  \href{http://dx.doi.org/10.1103/PhysRevD.101.064023}{{\em Phys. Rev. D}
  {\bfseries 101} no.~6, (2020) 064023},
  \href{http://arxiv.org/abs/1909.00096}{{\ttfamily arXiv:1909.00096 [gr-qc]}}.

\bibitem{Chakraborty:2016yna}
S.~Chakraborty, ``{Boundary Terms of the Einstein\textendash{}Hilbert
  Action},'' \href{http://dx.doi.org/10.1007/978-3-319-51700-1_5}{{\em Fundam.
  Theor. Phys.} {\bfseries 187} (2017) 43--59},
  \href{http://arxiv.org/abs/1607.05986}{{\ttfamily arXiv:1607.05986 [gr-qc]}}.

\bibitem{Parattu:2016trq}
K.~Parattu, S.~Chakraborty, and T.~Padmanabhan, ``{Variational Principle for
  Gravity with Null and Non-null boundaries: A Unified Boundary
  Counter-term},'' \href{http://dx.doi.org/10.1140/epjc/s10052-016-3979-y}{{\em
  Eur. Phys. J. C} {\bfseries 76} no.~3, (2016) 129},
  \href{http://arxiv.org/abs/1602.07546}{{\ttfamily arXiv:1602.07546 [gr-qc]}}.

\bibitem{Parattu:2015gga}
K.~Parattu, S.~Chakraborty, B.~R. Majhi, and T.~Padmanabhan, ``{A Boundary Term
  for the Gravitational Action with Null Boundaries},''
  \href{http://dx.doi.org/10.1007/s10714-016-2093-7}{{\em Gen. Rel. Grav.}
  {\bfseries 48} no.~7, (2016) 94},
  \href{http://arxiv.org/abs/1501.01053}{{\ttfamily arXiv:1501.01053 [gr-qc]}}.

\bibitem{Gupta:1993id}
K.~S. Gupta and S.~G. Rajeev, ``{Renormalization in quantum mechanics},''
  \href{http://dx.doi.org/10.1103/PhysRevD.48.5940}{{\em Phys. Rev. D}
  {\bfseries 48} (1993) 5940--5945},
  \href{http://arxiv.org/abs/hep-th/9305052}{{\ttfamily arXiv:hep-th/9305052}}.

\bibitem{BenAchour:2019ywl}
J.~Ben~Achour and E.~R. Livine, ``{Protected $SL(2,\mathbb{R})$ Symmetry in
  Quantum Cosmology},''
  \href{http://dx.doi.org/10.1088/1475-7516/2019/09/012}{{\em JCAP} {\bfseries
  09} (2019) 012}, \href{http://arxiv.org/abs/1904.06149}{{\ttfamily
  arXiv:1904.06149 [gr-qc]}}.

\bibitem{landau2013quantum}
L.~D. Landau and E.~M. Lifshitz, {\em Quantum mechanics: non-relativistic
  theory}, vol.~3.
\newblock Elsevier, 2013.

\bibitem{Feldbrugge:2017kzv}
J.~Feldbrugge, J.-L. Lehners, and N.~Turok, ``{Lorentzian Quantum Cosmology},''
  \href{http://dx.doi.org/10.1103/PhysRevD.95.103508}{{\em Phys. Rev. D}
  {\bfseries 95} no.~10, (2017) 103508},
  \href{http://arxiv.org/abs/1703.02076}{{\ttfamily arXiv:1703.02076
  [hep-th]}}.

\bibitem{FeldbruggeJobLeon2019}
{Feldbrugge, Job Leon}, {\em Path Integrals in the Sky: Classical and Quantum
  Problems with Minimal Assumptions}.
\newblock PhD thesis, 2019.
\newblock \url{http://hdl.handle.net/10012/15222}.

\bibitem{Hartle:1983ai}
J.~B. Hartle and S.~W. Hawking, ``Wave function of the universe,''
  \href{http://dx.doi.org/10.1103/PhysRevD.28.2960}{{\em Phys. Rev. D}
  {\bfseries 28} (Dec, 1983) 2960--2975}.
  \url{https://link.aps.org/doi/10.1103/PhysRevD.28.2960}.

\bibitem{Feldbrugge:2017fcc}
J.~Feldbrugge, J.-L. Lehners, and N.~Turok, ``{No smooth beginning for
  spacetime},'' \href{http://dx.doi.org/10.1103/PhysRevLett.119.171301}{{\em
  Phys. Rev. Lett.} {\bfseries 119} no.~17, (2017) 171301},
  \href{http://arxiv.org/abs/1705.00192}{{\ttfamily arXiv:1705.00192
  [hep-th]}}.

\bibitem{Feldbrugge:2017mbc}
J.~Feldbrugge, J.-L. Lehners, and N.~Turok, ``{No rescue for the no boundary
  proposal: Pointers to the future of quantum cosmology},''
  \href{http://dx.doi.org/10.1103/PhysRevD.97.023509}{{\em Phys. Rev. D}
  {\bfseries 97} no.~2, (2018) 023509},
  \href{http://arxiv.org/abs/1708.05104}{{\ttfamily arXiv:1708.05104
  [hep-th]}}.

\bibitem{Feldbrugge:2018gin}
J.~Feldbrugge, J.-L. Lehners, and N.~Turok, ``{Inconsistencies of the New
  No-Boundary Proposal},''
  \href{http://dx.doi.org/10.3390/universe4100100}{{\em Universe} {\bfseries 4}
  no.~10, (2018) 100}, \href{http://arxiv.org/abs/1805.01609}{{\ttfamily
  arXiv:1805.01609 [hep-th]}}.

\bibitem{Lehners2019}
A.~Di~Tucci, J.-L. Lehners, and L.~Sberna, ``No-boundary prescriptions in
  lorentzian quantum cosmology,''
  \href{http://dx.doi.org/10.1103/PhysRevD.100.123543}{{\em Phys. Rev. D}
  {\bfseries 100} (Dec, 2019) 123543}.
  \url{https://link.aps.org/doi/10.1103/PhysRevD.100.123543}.

\bibitem{olver2010nist}
F.~W. Olver, D.~W. Lozier, R.~F. Boisvert, and C.~W. Clark, {\em NIST handbook
  of mathematical functions hardback and CD-ROM}.
\newblock Cambridge university press, 2010.

\bibitem{Rajeev:2021zae}
K.~Rajeev, ``{A Lorentzian worldline path integral approach to Schwinger
  effect},'' \href{http://arxiv.org/abs/2105.12194}{{\ttfamily arXiv:2105.12194
  [hep-th]}}.

\bibitem{padmanabhan1981quantum}
T.~Padmanabhan, ``Quantum fluctuations and nonavoidance of the singularity in
  bianchi type i cosmology,'' {\em General Relativity and Gravitation}
  {\bfseries 13} no.~5, (1981) 451--455.

\bibitem{oyewumi2003isotropic}
K.~J. Oyewumi and E.~A. Bangudu, ``Isotropic harmonic oscillator plus inverse
  quadratic potential in n-dimensional spaces,'' {\em Arabian Journal for
  Science and Engineering} {\bfseries 28} no.~2, (2003) 173--182.

\bibitem{ChengandChan}
B.~K. Cheng and C.~F. T., ``An exact propagator for a time-dependent harmonic
  oscillator with a time-dependent inverse square potential,'' {\em Journal of
  Physica A: Mathematical and General} {\bfseries 20} no.~12, (1987) 3771.

\end{thebibliography}\endgroup

\bibliographystyle{utphys1}
\end{document}